%% file: main.tex
\documentclass[aps,twocolumn,superscriptaddress,amsmath,longbibliography,nofootinbib]{revtex4-1}
\usepackage[pdftex]{graphicx,color,hyperref,xcolor}
\usepackage{soul}
\usepackage{amsfonts}
\usepackage{amssymb}
\usepackage{amsmath}
\usepackage{amsthm}
\usepackage{bm}
\usepackage{braket}
\usepackage{enumerate}

\begin{document}

\title{Mott-moir\'e excitons}

\author{T.-S. Huang}
\affiliation{Joint Quantum Institute, University of Maryland, College Park, MD 20742, USA}

\author{Yang-Zhi Chou}
\affiliation{Condensed Matter Theory Center, University of Maryland, College Park, MD 20742, USA}
\affiliation{Joint Quantum Institute, University of Maryland, College Park, MD 20742, USA}

\author{C. L. Baldwin}
\affiliation{Joint Quantum Institute, University of Maryland, College Park, MD 20742, USA}

\author{Fengcheng Wu}
\affiliation{Key Laboratory of Artificial Micro- and Nano-structures of Ministry of Education, and School of Physics and Technology, Wuhan University, Wuhan 430072, China}
\affiliation{Wuhan Institute of Quantum Technology, Wuhan 430206, China}

\author{Mohammad Hafezi}
\affiliation{Joint Quantum Institute, University of Maryland, College Park, MD 20742, USA}
\affiliation{Pauli Center for Theoretical Studies, ETH Zurich, CH-8093 Zurich, Switzerland}

\date{\today}

\begin{abstract}
\input{section_0_Abstract}

\end{abstract}

\maketitle

\input{section_1_Intro}

\input{section_2_Summary}

\input{section_3_Formalism}

\input{section_4A_Single_holon_properties}

\input{section_4B_Exciton_properties}

\input{section_5_Conclusion}

\input{section_6_Acknowledgements}

\appendix

\input{section_A_dimensionless}

\input{section_A_HP}

\input{section_A_SCBA}

\input{section_A_HF}

\input{section_A_Mottmoire_current}

\input{section_A_moire_conductivity}

\input{section_A_Perturb}

\bibliography{Biblio_Mott_moire_Exciton}

\end{document}

%% file: section_0_Abstract.tex
We develop a systematic theory for excitons subject to Fermi-Hubbard physics in moir\'e twisted transition metal dichalcogenides (TMDs).
Specifically, we consider excitons in moir\'e systems for which the valence band is in the Mott-insulating regime.
These ``Mott-moir\'e excitons'', which are achievable in twisted TMD heterobilayers, are bound states of a magnetic polaron in the valence band and a free electron in the conduction band. 
We find significantly narrower exciton bandwidths in the presence of Hubbard physics, serving as a potential experimental signature of strong correlations.
We also demonstrate the high tunability of Mott-moir\'e excitons through the dependence of their binding energies, diameters, and bandwidths on the moir\'e period. 
Our work provides guidelines for future exploration of strongly correlated excitons in twisted TMD heterobilayers.


%% file: section_1_Intro.tex
\section{Introduction} \label{sec:introduction}

\begin{figure}[t]
\includegraphics[width=\columnwidth]{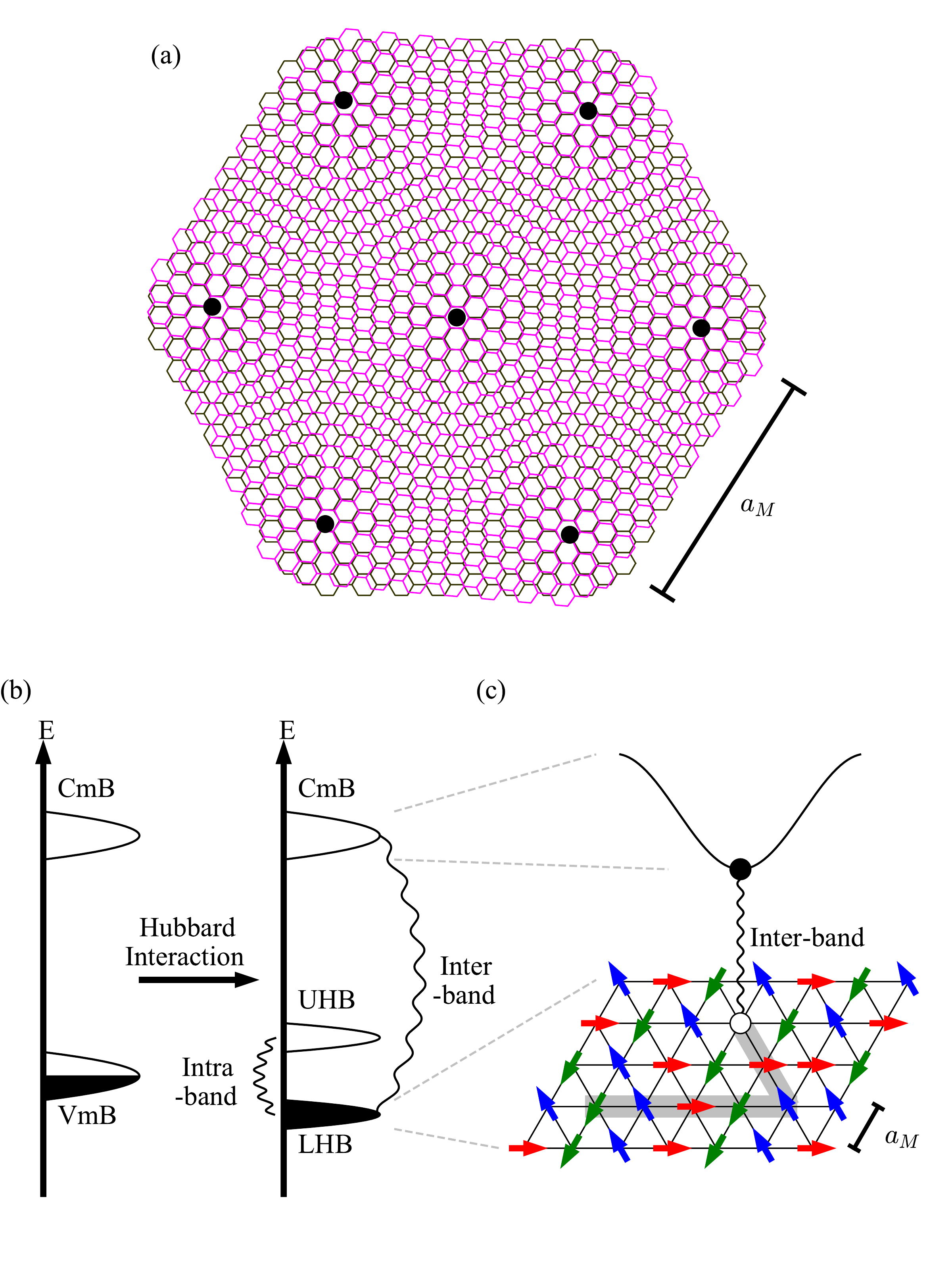}
\caption{Illustration of Mott-moir\'e excitons on the moir\'e super-lattice of a TMD heterobilayer. 
(a) Moir\'e super-lattice structure in a twisted TMD heterobilayer.
Black dots indicate AA-stacked atoms which themselves define a triangular lattice structure and lead to folded moir\'e bands.
(b) Schematic diagram of the valence (VmB) and conduction (CmB) moir\'e bands. 
Vertical axes indicate energy, and the shaded area refers to the filled Fermi sea. 
The Hubbard interaction within the VmB causes it to split into upper and lower Hubbard bands (UHB and LHB). 
Black wavy lines indicate interactions that can form Mott excitons, either intra-band (hole in LHB and electron in UHB~\cite{Huang2020Spin}) or inter-band (hole in LHB and electron in CmB).
The latter is the subject of this paper.
(c) Schematic diagram for the inter-band Mott-moir\'e exciton. 
Black and white dots indicate the CmB electron and VmB hole respectively.
Red, green, and blue arrows show the 120$^\circ$ spin-ordered state on the triangular super-lattice in the VmB.
Note that the trajectory of the hole, represented by gray shading, 
displaces spins and thus disturbs the spin order.
}
\label{fig_Mott_moire_exciton_illustrate}
\end{figure}

Two-dimensional (2D) semiconducting transition metal dichalcogenides (TMDs)~\cite{Kang2013Band,Xiao2012,Stier2018Magnetooptics,Chernikov2014Exciton,Wu2015Exciton,Onga2017exciton,Goryca2019revealing,Selig2016excitonic,Moody2015intrinsic,Hao2016direct,Li2020Spontaneous,Efimkin2017Many,Efimkin2018Exciton,Smolenski2019Interaction,Back2018Realization,Zhang2017Interlayer,Wu2018Theory,Wu2017Topological,Naik2022Nature,Rivera2018Interlayer,Seyler2019Signatures,Tran2019Evidence,Alexeev2019Resonantly,Wu2018Hubbard,Wu2019Topological,Pan2020Band,Pan2020quantum,Zang2021HF,Hu2021Competing,Zhang2021Electronic,Tang2020Simulation,Regan2020Mott,Xu2020Correlated,Li2021Continuous,Li2021Imaging,Campbell2022Strongly,Regan2020Mott,Jin2021Stripe,Xu2022tunable,Choi2020Moire,Li2020Exciton} have become a rich platform with which to explore the interplay of optoelectronics and many-body physics, due primarily to their band structure properties such as infrared/visible-frequency band gaps~\cite{Kang2013Band} and additional valley degrees of freedom at low energy~\cite{Xiao2012}.
In particular, many studies have focused on the properties of excitons (bound states of electrons and holes)~\cite{Stier2018Magnetooptics,Chernikov2014Exciton,Wu2015Exciton,Onga2017exciton,Goryca2019revealing,Selig2016excitonic,Moody2015intrinsic,Hao2016direct}, and on understanding how excitons interact with the Fermi sea to form exciton-polarons~\cite{Li2020Spontaneous,Efimkin2017Many,Efimkin2018Exciton,Smolenski2019Interaction,sidler2017fermi}. 
In TMD \textit{bilayers} [Fig.~\ref{fig_Mott_moire_exciton_illustrate}(a)], the relative twist angle between the two layers, and the resulting super-lattice period, is a further tunable parameter~\cite{Wu2018Theory,Wu2017Topological,Naik2022Nature,Rivera2018Interlayer,Seyler2019Signatures,Tran2019Evidence,Alexeev2019Resonantly}.
The electronic properties of twisted TMD bilayers are very different from those of monolayers due to the presence of flat moir\'e bands that significantly enhance the role of many-body interactions, leading to strong correlations~\cite{Wu2018Hubbard,Wu2019Topological,Pan2020Band,Pan2020quantum,Zang2021HF,Hu2021Competing,Zhang2021Electronic,Tang2020Simulation,Regan2020Mott,Xu2020Correlated,Li2021Continuous,Li2021Imaging,Campbell2022Strongly,Regan2020Mott,Jin2021Stripe,Xu2022tunable}.
The effect of strong correlations on excitons in the presence of moir\'e structure remains a subject of active investigation. 

One consequence of strong-correlation physics in twisted TMD bilayers is the emergence of correlated insulating states and charge order~\cite{Wu2018Hubbard,Wu2019Topological,Pan2020Band,Pan2020quantum,Zang2021HF,Hu2021Competing,Zhang2021Electronic,Tang2020Simulation,Regan2020Mott,Xu2020Correlated,Li2021Continuous,Li2021Imaging,Campbell2022Strongly,Regan2020Mott,Jin2021Stripe,Xu2022tunable}.
It has been pointed out that generalized moir\'e-Hubbard models can emerge for the first valence moir\'e band (VmB) in heterobilayers~\cite{Wu2018Hubbard} and for the first few VmBs in homobilayers~\cite{Wu2019Topological,Pan2020Band,Zhang2021Electronic}.
Moreover, at certain filling fractions $\nu$ (i.e., number of electrons per super-lattice unit cell), these models predict the existence of correlated states such as Mott insulators ($\nu=1$) and Wigner crystals ($\nu=\frac{1}{4},\frac{1}{3},\frac{1}{2},\frac{2}{3},\frac{3}{4}$)~\cite{Pan2020quantum,Zhang2021Electronic}.
This explains several transport~\cite{Tang2020Simulation,Regan2020Mott} and optical~\cite{Tang2020Simulation,Regan2020Mott,Campbell2022Strongly,Xu2020Correlated} measurements of TMD heterobilayers, which observe enhanced resistivity and incompressibility at the aforementioned filling fractions.

In addition to charge order, spin order~\cite{Zang2021HF,Hu2021Competing} can significantly influence the properties of twisted TMD bilayers.
For example, at half filling, a triangular-lattice Hubbard model yields a 120$^{\circ}$ magnetically-ordered state~\cite{Wu2018Hubbard,Azzouz1996Motion,Chen2022proposal}, and the spin fluctuations on top of such a background can strongly renormalize the charge dynamics, giving rise to magnetic polarons~\cite{Martinez1991Spin,Vojta1999,Grusdt2018Parton}.
Intuitively, this is because the movement of charges in the ordered state disturbs the spin configuration, leaving a trail of misaligned spins that is energetically unfavored. 
To the best of our knowledge, conclusive signatures of spin ordering and magnetic polarons in twisted TMD bilayers have not been established experimentally, nor has the question been answered of how spin order affects the excitons.

The rich phenomena derived from the Hubbard model motivates us to study ``Mott excitons'' in twisted TMD bilayers, namely excitons in which one or both of the charge constituents are magnetic polarons rather than bare charges.
Broadly speaking, two distinct types of Mott excitons can exist.
We coin them ``intra-band'' and ``inter-band'' Mott excitons [see Fig.~\ref{fig_Mott_moire_exciton_illustrate}(b)].

Intra-band Mott excitons consist of a vacancy and a double-occupancy within a single-band Hubbard model.
Since the constituent charges lie within the same Bloch band, the lowest such excitonic state is optically dark (within the dipole approximation) and therefore not readily accessible in solid-state systems.
A few theoretical works have considered this type of exciton~\cite{Essler2001Excitons,Wrobel2002Excitons,Jeckelmann2003Optical,Huang2020Spin}, with particular focus on how spin fluctuations provide the binding mechanism~\cite{Huang2020Spin}, and despite the experimental challenges, certain indirect optical signatures of intra-band Mott excitons have recently been reported in iridates~\cite{Alpichshev2015Confinement,Alpichshev2017Origin}.

On the other hand, inter-band Mott excitons consist of a vacancy and electron in separate bands, with the valence band described by a Hubbard model and the conduction band otherwise empty.
In this case, the binding mechanism has a direct Coulomb origin rather than being spin-mediated.
These excitons very well can be optically bright, assuming the valence and conduction bands satisfy the appropriate selection rule. 
Accordingly, inter-band Mott excitons have recently been reported in cuprates via reflectivity measurements~\cite{Terashige2019Doublon}. 

In this paper, we investigate the \textit{inter}-band Mott exciton formed from a magnetic polaron in the valence moir\'e band (VmB) and an electron in the conduction moir\'e band (CmB).
We refer to these throughout as ``Mott-moir\'e excitons''.
We give a theoretical description for the formation of Mott-moir\'e excitons, and identify the role of spin-ordering in determining their properties.
In particular, we compare Mott-moir\'e excitons to those that would exist in the same band structure with the same Coulomb interaction but without any Mott physics (we label the latter simply as ``moir\'e excitons'').
Finally, many of our techniques and conclusions hold equally well for Mott excitons in non-moir\'e systems.

The outline of the paper is as follows.  
We summarize the model and our main results in Sec.~\ref{sec:summary}.
We describe our theoretical techniques in Sec.~\ref{sec:formalism}, and present our results in more detail in Sec.~\ref{sec:results}.
Finally, we discuss potential experimental signatures of Mott-moir\'e excitons in Sec.~\ref{sec:conclusion}.
Various technical details can be found in the appendices.


%% file: section_2_Summary.tex
\section{Summary} \label{sec:summary}

\subsection{Overview of the model}
We consider Mott excitons in the presence of a moir\'e potential coming from a twisted TMD heterobilayer system [see Fig.~\ref{fig_Mott_moire_exciton_illustrate}(a)].
Stacking the two monolayers with a small relative twist angle gives the sample a moir\'e period $a_M$ greater than the monolayer lattice spacings.
This enlarged periodicity folds the band structure into moir\'e bands.
It is known that a tight-binding model in terms of super-lattice sites can describe both the lowest CmB and the highest VmB, albeit with strong on-site interactions in the latter at half filling~\cite{Li2021Imaging,Wu2018Hubbard}.
Hence, we focus on a two-moir\'e-band model to capture the essence of inter-band Mott excitons:
\begin{equation}
\label{eq:Starting_model}
\begin{aligned}
\hat{H}
&=
-t
\sum_{\tau}
\sum_{\langle\bm{R},\bm{R}'\rangle}
\left[
\hat{c}_{\bm{R},\tau}^{\dag}
\hat{c}_{\bm{R}',\tau}
+
\hat{h}_{\bm{R},\tau}^{\dag}
\hat{h}_{\bm{R}',\tau}
\right]
\\
&
\quad +
U\sum_{\bm{R}}
\hat{n}_{\bm{R},\uparrow}
\hat{n}_{\bm{R},\downarrow}
\\
&
\quad -
\sum_{\tau\tau'}
\sum_{\bm{R}\bm{R}'}
V_{|\bm{R}-\bm{R}'|}
\hat{c}_{\bm{R},\tau}^{\dag}
\hat{h}_{\bm{R}',\tau'}^{\dag}
\hat{h}_{\bm{R}',\tau'}
\hat{c}_{\bm{R},\tau}
,
\end{aligned}
\end{equation}
with $\tau \in \{ \uparrow, \downarrow \}$ labeling the valley index (equivalently spin index --- the two are locked together in TMDs~\cite{Rivera2018Interlayer}), and $\langle \bm{R},\bm{R}' \rangle$ denoting nearest-neighbor sites on a triangular super-lattice.
$\hat{c}_{\bm{R},\tau}$ represents the CmB electron annihilation operator and $\hat{h}_{\bm{R},\tau}$ the VmB hole operator.
We assume that the charges live on the same triangular super-lattice, although they could lie on different lattices in reality~\cite{Naik2022Nature}.
$\hat{n}_{\bm{R},\tau}\equiv1-\hat{h}_{\bm{R},\tau}^{\dag}\hat{h}_{\bm{R},\tau}$ is the \textit{electron} occupation at moir\'e site $\bm{R}$ and valley $\tau$ in the VmB.
Note that, as discussed above, we only include the Hubbard interaction $U$ for electrons in the VmB, specifically on-site repulsion since the off-site electrostatic interactions can be rendered insignificant by gate-screening~\cite{Wu2018Hubbard}.
We choose the VmB and CmB hopping coefficients $t$ to be equal for simplicity, and assume $U\gg t$~\cite{Wu2018Hubbard}.
Lastly, $V_{|\bm{R}-\bm{R}'|}$ denotes the Coulomb interaction between the two moir\'e bands.
Since the VmB is in a (correlated) insulating state and the CmB is initially empty, the interaction is not screened.

One may wonder whether the strong Coulomb binding in excitons, which is larger than the moir\'e bandwidths, renders the single-particle bands irrelevant.
However, experimental observations suggest that the incompressibility of Mott states in the single-particle bands does manifest as a modification of the exciton energy nonetheless~\cite{Tang2020Simulation,Regan2020Mott,Campbell2022Strongly,Xu2020Correlated}.   
Thus we expect the above two-band model to capture the relevant strong-correlation physics. 
Note that this assumption is even more justified for excitons with higher principal quantum numbers (and hence smaller binding energies).

Following the standard arguments~\cite{Auerbach1994}, including extra charges (in our case vacancies) into the half-filled VmB yields an effective t-J model. 
We take the 120$^\circ$ coplanar spin-ordered phase of such a model as our ground state~\cite{Wu2018Hubbard}, but still include spin fluctuations.
These propagate at energy scale $J\simeq4t^2/U$~\cite{Wu2018Hubbard} and dress the charges into magnetic polarons~\cite{Martinez1991Spin,Vojta1999,Grusdt2018Parton}. 
It is convenient to describe the charge and spin degrees of freedom separately, via slave fermion~\cite{Han2016Charge,Martinez1991Spin} and Holstein-Primakoff bosons~\cite{Auerbach1994} respectively.
This ultimately (see Sec.~\ref{sec:formalism}) reduces Eq.~\eqref{eq:Starting_model} to the following two-body Hamiltonian:
\begin{equation} \label{eq:two_body_Hamiltonian}
\begin{aligned}
\hat{H}
&=
\sum_{\bm{k}}
\epsilon_{\bm{k}}
\hat{\psi}_{\bm{k}}^\dag
\hat{\psi}_{\bm{k}}
-
2t
\sum_{\bm{k},\tau}
\gamma_{\bm{k}}
\hat{c}_{\bm{k},\tau}^\dag
\hat{c}_{\bm{k},\tau}
\\
&
\quad -
\frac{1}{{\cal{A}}}
\sum_{\tau}
\sum_{\bm{k},\bm{k}',\bm{q}}
V(q)
\hat{c}_{\bm{k}+\bm{q},\tau}^\dag
\hat{\psi}_{\bm{k}'-\bm{q}}^\dag
\hat{\psi}_{\bm{k}'}
\hat{c}_{\bm{k},\tau}
,
\end{aligned}
\end{equation}
in which $\hat{\psi}$ stands for the fermionic charge degree of freedom (i.e., holon) in the VmB, whereas $\hat{c}$ remains the bare CmB electron.
${\cal{A}}$ denotes the system area.
Momentum sums run over the first moir\'e Brillouin zone (mBZ).
$\epsilon_{\bm{k}}$ is the (dressed) holon dispersion, $-2t\gamma_{\bm{k}}$ is the CmB electron dispersion, and $V(q)$ is the Coulomb interaction written in momentum space.
See Eqs.~\eqref{eq:approximate_single_particle_dispersion},~\eqref{eq:gamma_coefficient}, and~\eqref{eq:Coulomb_gate_screening} for the explicit expressions and further details.

Eq.~\eqref{eq:two_body_Hamiltonian} captures the formation of Mott-moir\'e excitons from electrons in the CmB and holons in the VmB [see also Fig.~\ref{fig_Mott_moire_exciton_illustrate}(b)].
We introduce the composite boson operator $\hat{X}_{n,\tau}(\bm{Q})$ for such a bound state, which we write in the form
\begin{equation}
\label{eq:Exciton_operator}
\hat{X}_{n,\tau}(\bm{Q})
=
\sum_{\bm{p}}
\phi_{\bm{Q}}^{(n)}(\bm{p})
\hat{\psi}_{\frac{\bm{Q}}{2}-\bm{p}}
\hat{c}_{\frac{\bm{Q}}{2}+\bm{p},\tau}
,
\end{equation}
where $\bm{Q}$ and $\bm{p}$ are the total and relative momenta of the two particles respectively, and $n$ labels the internal state of the exciton.
$\phi_{\bm{Q}}^{(n)}(\bm{p})$ is the wavefunction of the exciton, and if chosen so as to solve an appropriate effective two-particle Schrodinger equation (Eq.~\eqref{eq:Wannier_equation}), Eq.~\eqref{eq:two_body_Hamiltonian} becomes ``quadratic'' in terms of the composite boson operators:
\begin{equation}
\label{eq:Exciton_Hamiltonian}
\hat{H}
=
\sum_{\bm{Q},\tau}
E_{n,\bm{Q}}^X
\hat{X}_{n,\tau}^\dag(\bm{Q})
\hat{X}_{n,\tau}(\bm{Q})
,
\end{equation}
with $E_{n,\bm{Q}}^X$ denoting the exciton energy.
The operator $\hat{X}_{n,\tau}(\bm{Q})$ can be shown to satisfy bosonic commutation relations in the dilute limit~\cite{Haug1984Electron}, meaning that Eq.~\eqref{eq:Exciton_Hamiltonian} does amount to an approximate diagonalization of the Hamiltonian for small numbers of excitons.

\begin{figure}[t]
\centering
\includegraphics[width=1.0\columnwidth]{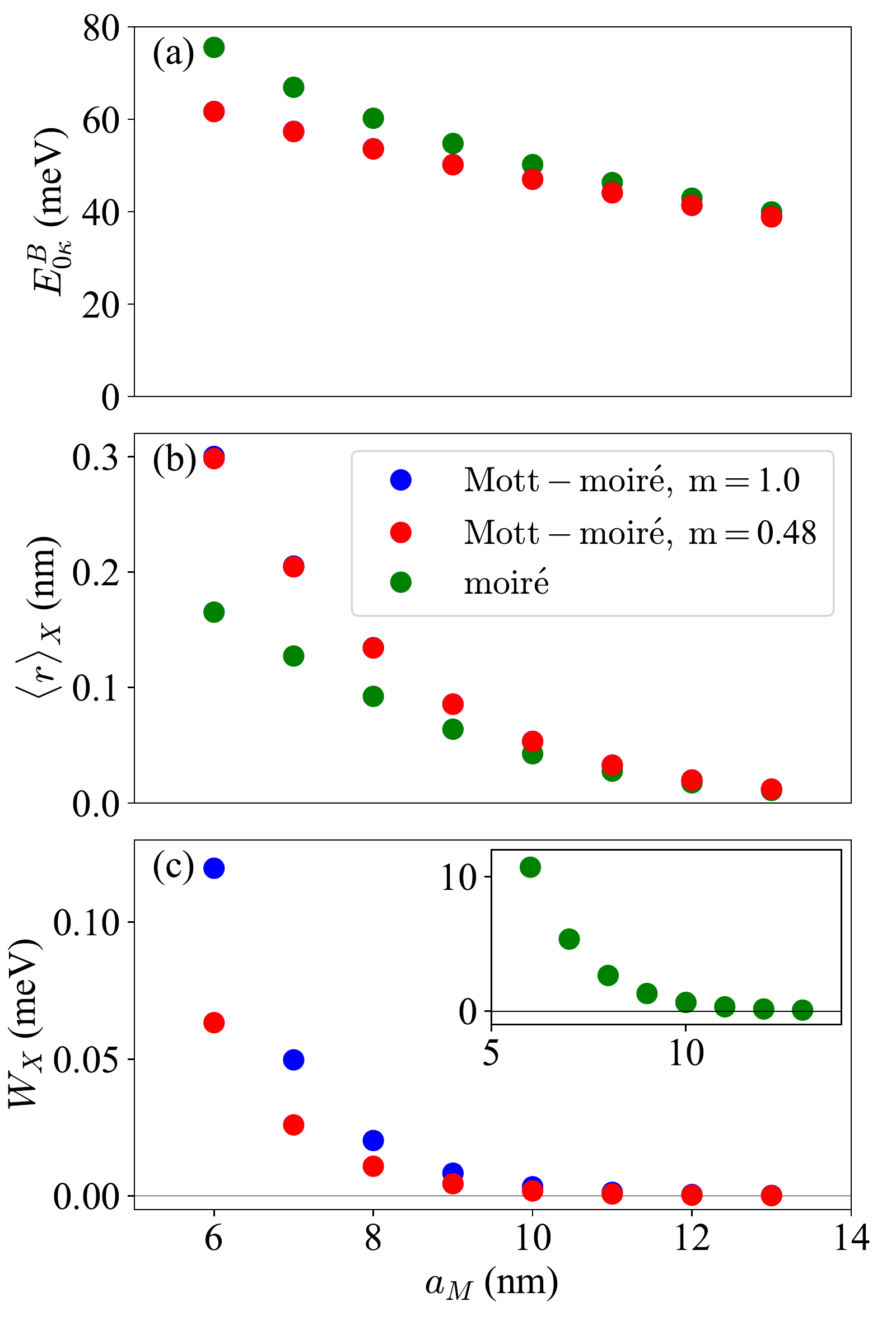}
\caption{Properties of Mott-moir\'e excitons at different magnetization (blue and red, indistinguishable at the scales of top and middle panels) and moir\'e excitons (green) as functions of the moir\'e period $a_M$. 
Dielectric constant is $\epsilon_r=10$.
System size is $N=3\times24^2$ sites.
(a) Binding energy of the lowest internal state $E_{0,\kappa}^B$ at total momentum $\bm{Q}=\bm{\kappa}$, which has the largest binding among all $\bm{Q}$ for both excitons (even though the moir\'e exciton energy is lower at $\bm{Q} = \Gamma$ in absolute numbers).
(b) Average diameter of excitons at total momentum $\kappa$. 
(c) Exciton bandwidths $W_X$. 
Inset shares the same axes.
Values for $t$ and $J$ as functions of $a_M$ are taken from Ref.~\cite{Wu2018Hubbard} for WSe\textsubscript{2} on top of MoSe\textsubscript{2} (see also Fig.~\ref{fig_Jandt}). 
}
\label{fig_exciton_properties}
\end{figure}

We end this overview by noting that, strictly speaking, Eq.~\eqref{eq:Exciton_operator} gives the Mott-moir\'e exciton as a composite particle involving a \textit{holon} rather than magnetic polaron.
The holon is merely the charge sector of the polaron --- the latter additionally contains a surrounding cloud of spin fluctuations~\cite{Grusdt2019Microscopic,Martinez1991Spin}.
Yet since we shall find that the exciton radius (Fig.~\ref{fig_exciton_properties}) is smaller than the polaron radius~\cite{Grusdt2019Microscopic}, we feel it is reasonable to consider binding between the electron and holon alone (spin fluctuations are still included via the dressed holon dispersion).


\begin{figure}[t]
\centering
\includegraphics[width=1.0\columnwidth]{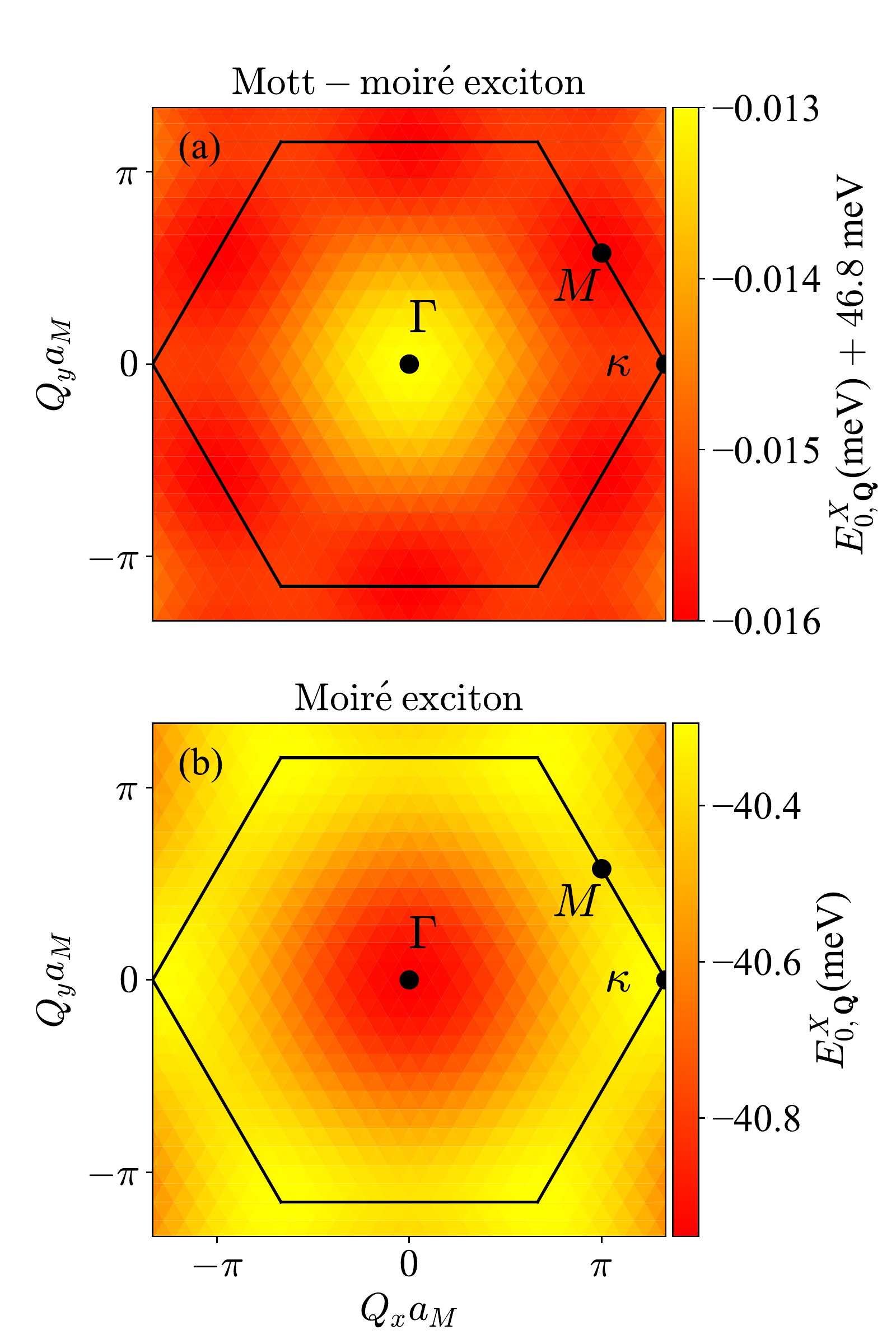}
\caption{Dispersion relation of the lowest-energy throughout the entire mBZ, indicated by the black hexagon for Mott-moir\'e (top) and moir\'e (bottom) exciton, at super-lattice period $a_M = 10$ nm and with dielectric constant $\epsilon_r = 10$.
For Mott-moir\'e, we use $t/J = 7.3$ (taken from Ref.~\cite{Wu2018Hubbard}) and the equilibrium magnetization $m = 0.48$.
System size is $3\times 24^2$ sites.
$Q_x$ and $Q_y$ are the total momentum of the two-particle state.
Colorbars indicate energy relative to the two-particle continuum (lowest energy of two free particles) --- note in particular that the top panel has energies shifted by $46.8$ meV.
Blue dots indicate important points in the Brillouin zone.
}
\label{fig_exciton_dispersion_colormap}
\end{figure}

\begin{figure}[t]
\centering
\includegraphics[width=1.0\columnwidth]{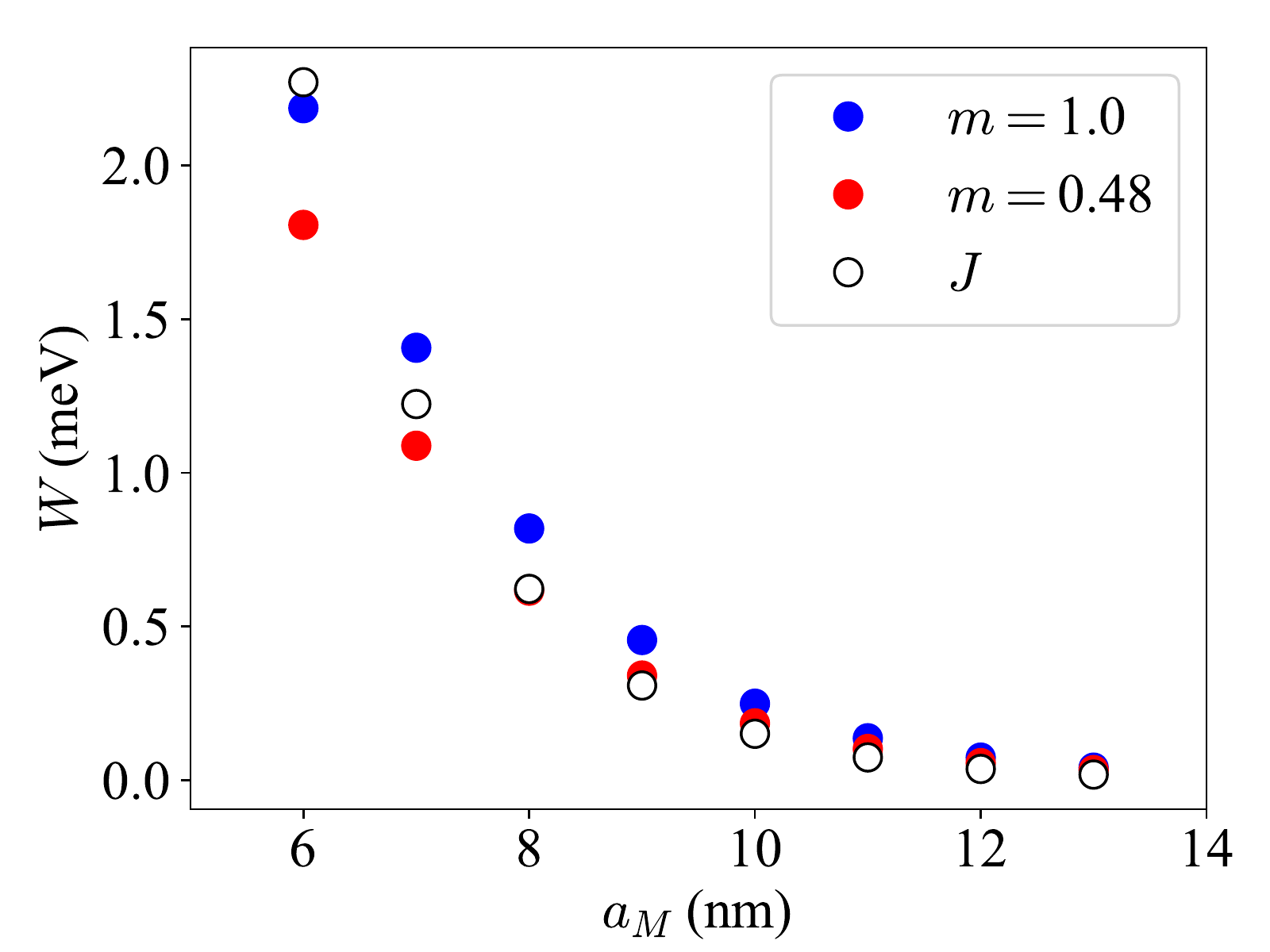}
\caption{Dressed holon bandwidth $W$ from SCBA as a function of moir\'e period $a_M$, at different sublattice magnetizations $m$ (blue and red).
System size is $3\times24^2$ sites.
Also shown is $J$ as a function of $a_M$ (empty circle), for WSe\textsubscript{2} on top of MoSe\textsubscript{2} according to Ref.~\cite{Wu2018Hubbard}.}
\label{fig_holon_bandwidth}
\end{figure}

\subsection{Overview of results} Our main finding is that moir\'e and Mott-moir\'e excitons are similar in certain regards (namely the binding energy and radius) but dramatically different in others (particularly the bandwidth, i.e., exciton mass). 
We further identify how the properties of the two vary with the moir\'e period $a_M$ --- recall that the moir\'e period is tunable experimentally.

To begin, the dispersions of moir\'e and Mott-moir\'e excitons are quite different, as shown in Fig.~\ref{fig_exciton_dispersion_colormap}.
Whereas moir\'e excitons possess a band minimum at $\bm{Q} = \Gamma \equiv (0, 0)$ and maxima at the mBZ boundary, Mott-moir\'e excitons have the opposite behavior: a maximum at $\bm{Q} = \Gamma$ and minima at the mBZ boundary.
We shall demonstrate that the inverted dispersion is precisely a consequence of the background spin order.
Furthermore, the bandwidth $W_X$ of Mott-moir\'e excitons is roughly two orders of magnitude smaller than that of moir\'e excitons [see Fig.~\ref{fig_exciton_properties}(c)]. 
This suppression is primarily due to the reduced holon bandwidth, and experiments in cold-atom quantum simulators have reported similar effects~\cite{ji2021coupling}.
Interestingly, in a sense we shall make sharp, the lowered holon bandwidth is more a consequence of spin \textit{fluctuations} than spin order alone.
We refer to Section~\ref{sec:results} for more details.

Mott-moir\'e excitons have a slightly smaller binding energy $E_{n,\bm{Q}}^B$ and larger diameter $\langle r\rangle_X$ compared to moir\'e excitons in their lowest internal states ($n=0$) [see Fig.~\ref{fig_exciton_properties}(a) and (b)]. 
Regardless, in both cases the exciton is significantly smaller than a moir\'e period, and correspondingly the binding energy is much greater than the Coulomb energy scale for charges separated by $a_M$.
Qualitatively, this is due to the fact that the on-site Coulomb attraction is noticeably larger than the super-lattice hopping amplitudes (see Fig.~\ref{fig_kinetic_potential}). 
Thus these bound states are of the Frenkel variety~\cite{Agranovich1968Collective}, and we give a corresponding analysis in Sec.~\ref{sec:results}.
This is quite different from the Wannier regime found in conventional semiconductors, for which the excitons are larger than the lattice scales of the problem~\cite{Haug2004}.

As for how these properties vary with the moir\'e period $a_M$, a larger period implies significantly suppressed super-lattice hopping amplitudes and thus relatively stronger Coulomb binding.
This explains the trends seen in Fig.~\ref{fig_exciton_properties}: $\langle r\rangle_X$ and $W_X$ both decrease as $a_M$ increases.
Since the inter-site Coulomb interaction is itself weaker at larger $a_M$, albeit less so than the hopping strength, the binding energy $E_{0 \kappa}^B$ decreases as well. 
We refer to Sec.~\ref{sec:results} and Appendix~\ref{Appendix_perturbation} for more details.

We also compare the Mott-moir\'e exciton properties at different sublattice magnetizations $m$ (the order parameter for the 120$^\circ$ coplanar spin state).
As we are considering 2D systems, spin fluctuations reduce the magnetization even at zero temperature.
Linear spin-wave theory predicts $m \approx 0.48$ on the triangular lattice, which we compare to full magnetization $m = 1$. 
The qualitative trends for all properties are the same at both magnetizations.
Furthermore, we see in Fig.~\ref{fig_exciton_properties} that only the exciton bandwidth has a noticeable dependence on $m$ (and even then only by a factor of 2).
This is because $m$ influences only the holon kinetic energy, which is a small energy scale regardless.
Thus while the exciton bandwidth (being controlled primarily by the holon bandwidth) is sensitive to magnetization, the other properties (for which the holon acts more-or-less as inert) are not.

Lastly, we study the excited states of Mott-moir\'e excitons from the two-band model Eq.~\eqref{eq:two_body_Hamiltonian}.
In accordance with the symmetry group of this model, we identify states that can be classified as s-, p-, d-, and f-wave.
However, we find that only s-wave excitons are optically bright (see Eq.~\eqref{eq:Optical_conductivity_Mott}) and that the oscillator strength comes mainly from the lowest state (see Fig.~\ref{fig_exciton_level}).
Although these results are based on a two-band model, we expect that the analysis can be generalized to multi-band models.

%% file: section_3_Formalism.tex
\section{Formalism and methods} \label{sec:formalism}

In this section, we present the formalism describing inter-band Mott-moir\'e excitons in TMD heterobilayers. 
The Hubbard model on a triangular lattice has been investigated with various analytical methods: Hartree-Fock mean field theory~\cite{Fujita1992Triple}, strong-coupling expansions~\cite{Yang2010Effective}, and slave particles~\cite{Azzouz1996Motion,Chen2022proposal}.
Here we use the slave-particle formalism to study the dressed holon because spin and charge excitations are automatically distinguished in this approach.
The steps of our calculations are summarized as follows:
\begin{enumerate}[i)]
    \item Implement projection to the subspace of zero double-occupancies in the VmB and keep only nearest-neighbor terms, thus obtaining a t-J model~\cite{Auerbach1994,Wu2018Hubbard}.

    \item Express the Hamiltonian in terms of slave particles, namely holons and spinons (keep in mind that the spin degrees of freedom described by spinons are locked to the valley degrees of freedom).
    
    \item Focus on the 120$^{\circ}$ coplanar magnetically ordered phase of the triangular-lattice t-J model, as described through a mean-field approximation for the spinons (while still including linear spin-wave fluctuations).
    
    \item Calculate the dispersion of spin-dressed holons within the self-consistent Born approximation (SCBA)~\cite{Azzouz1996Motion,Chen2022proposal,Martinez1991Spin,Han2016Charge}.
    
    \item Construct the exciton Hamiltonian from the kinetic energies of dressed holons in the VmB and electrons in the CmB, together with the Coulomb interaction. Diagonalize this Hamiltonian numerically to obtain the exciton spectrum and wavefunctions.
\end{enumerate}

Before proceeding, let us emphasize that our usage of mean-field theory to describe the magnetic order implies that our results become inaccurate near its melting point.
We nonetheless expect mean-field theory to capture the qualitative features of the 120$^\circ$ spin-ordered phase, and previous studies have confirmed that magnetic order persists (at around 0.4 -- 0.5 of the classical value) even once quantum fluctuations are taken into account~\cite{White2007Neel,Capriotti1999Long,Jolicoeur1989Spin}.
Furthermore, our results turn out to be largely insensitive to the precise value of the magnetization (see Sec.~\ref{sec:summary}).

\subsection{t-J model}

Since the derivation of a t-J model from a half-filled Hubbard model is by now standard (see, e.g., Ref.~\cite{Auerbach1994}), we simply mention the result.
Starting from a state with one VmB electron per super-lattice site, second-order perturbation theory in $t/U$ gives an effective Hamiltonian
\begin{equation} \label{eq:tJ_model_def}
\hat{H}_{\textrm{tJ}}
=
-t
\sum_{\tau}
\sum_{\langle \bm{R},\bm{R}' \rangle}
\hat{\cal{P}}
\hat{h}_{\bm{R},\tau}^{\dag}
\hat{h}_{\bm{R}',\tau}
\hat{\cal{P}}
+
J\sum_{\langle \bm{R},\bm{R}' \rangle}
\hat{\bm{S}}_{\bm{R}}
\cdot
\hat{\bm{S}}_{\bm{R}'},
\end{equation}
where $J \equiv 4t^2/U$, $\hat{\cal{P}}$ is the projector onto the subspace having no more than one electron per site, and $\hat{\bm{S}}_{\bm{R}} \equiv \frac{1}{2}\sum_{\tau\tau'}\hat{h}_{\bm{R},\tau} \bm{\sigma}_{\tau,\tau'}\hat{h}_{\bm{R},\tau'}^{\dag}$ with
$\bm{\sigma}_{\tau,\tau'}$ denoting the vector of 2 $\times$ 2 Pauli matrices.

\subsection{Slave particles}

\begin{figure}[t]
\centering
\includegraphics[width=\columnwidth]{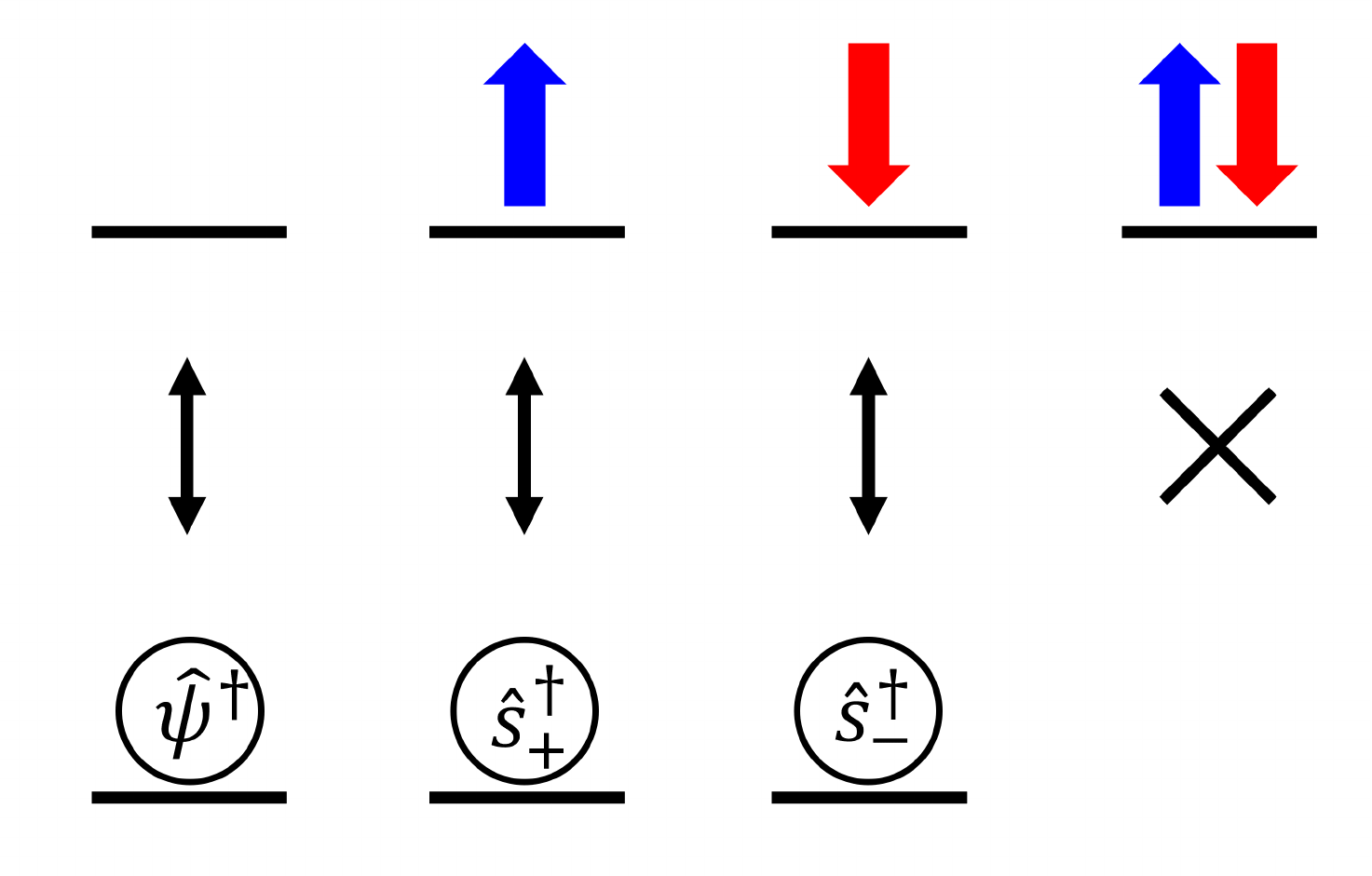}
\caption{Schematic diagram for the slave-particle formalism. 
Blue and red arrows represent the states with a single $\tau = \uparrow$ and $\tau = \downarrow$ electron respectively, and are mapped to states with the corresponding bosonic spinons $\hat{s}$. 
States with zero and two electrons are mapped to corresponding fermionic slave particles, holons $\hat{\psi}$ and doublons $\hat{d}$. 
Doublons are not shown, as indicated by the cross, because they are projected out due to the large energy cost $U$.}
\label{fig_slave_particle}
\end{figure}

The hole creation operator can be represented as (meaning that the two sides obey the same commutation relations)
\begin{equation} \label{eq:slave_particle_def}
\hat{h}_{\bm{R},\tau}^\dagger
=
\hat{\psi}_{\bm{R}}^{\dag}
\hat{s}_{\bm{R},\tau} + \tau \hat{s}_{\bm{R}, -\tau}^{\dag} \hat{d}_{\bm{R}},
\end{equation}
with fermionic $\hat{\psi}_{\bm{R}}$ and $\hat{d}_{\bm{R}}$, and bosonic $\hat{s}_{\bm{R},\tau}$.
We interpret $\hat{\psi}_{\bm{R}}$ as an empty site --- a ``holon'' --- and $\hat{s}_{\bm{R},\tau}$ as a singly-occupied site with spin (equivalently valley) --- a ``spinon''~\cite{Frohlich1992Slave,Feng1994Fermion,Shimizu2011Antiferromagnetic}.
$\hat{d}_{\bm{R}}$ corresponds to a doubly-occupied site, but since the t-J model projects into the subspace with no double occupancies, this operator does not appear in any subsequent expressions (it is needed only to ensure that Eq.~\eqref{eq:slave_particle_def} is consistent with the commutation relations).
The slave-particle transformation is illustrated in Fig.~\ref{fig_slave_particle}.

Eq.~\eqref{eq:slave_particle_def} indicates that hole creation (i.e., electron annihilation) is equivalent to removing the corresponding spinon and creating a holon in its place (or replacing a double occupancy with the non-annihilated spin).
Since we are neglecting double occupancies, any site which does not contain a spin by definition contains a hole, and therefore the slave particles must obey the following constraints for all $\bm{R}$:
\begin{equation} \label{eq:slave_particle_constraint}
\hat{\psi}_{\bm{R}}^{\dag}\hat{\psi}_{\bm{R}}
+
\sum_{\tau} \hat{s}_{\bm{R} \tau}^{\dag}\hat{s}_{\bm{R} \tau}
= 1, \qquad \hat{d}_{\bm{R}}^{\dag} \hat{d}_{\bm{R}} = 0.
\end{equation}
Substituting Eq.~\eqref{eq:slave_particle_def} into Eq.~\eqref{eq:tJ_model_def} and making use of the constraints allows us to express $\hat{H}_{\textrm{tJ}}$ as $\hat{H}_{t}+\hat{H}_{J}$, where:
\begin{equation} \label{eq:slave_particle_H_t}
\hat{H}_{t}
=
-t
\sum_{\tau}
\sum_{\langle \bm{R},\bm{R}' \rangle} \big(
\hat{\psi}_{\bm{R}}^{\dag} 
\hat{\psi}_{\bm{R}'}
\hat{s}_{\bm{R}',\tau}^{\dag}
\hat{s}_{\bm{R},\tau}
+
h.c.
\big)
,
\end{equation}
\begin{equation}
\label{eq:slave_particle_H_J}
\hat{H}_{J}
=
J\sum_{\langle \bm{R},\bm{R}' \rangle}
\hat{\bm{{\cal{S}}}}_{\bm{R}}
\cdot
\hat{\bm{{\cal{S}}}}_{\bm{R}'}
,
\end{equation}
where the spin vectors are now expressed in terms of spinons: $\hat{\bm{{\cal{S}}}}_i=\frac{1}{2}\sum_{\tau\tau'}\hat{s}_{\bm{R},\tau}^{\dag}\hat{\bm{\sigma}}_{\tau,\tau'}\hat{s}_{\bm{R},\tau'}$. 
Note that the Hamiltonian automatically preserves the conditions in Eq.~\eqref{eq:slave_particle_constraint}.

\subsection{Magnetic order and spin waves}

To study the magnetic order in the t-J model, we consider the dilute limit in which the low number of holons does not disturb the spin background.
Consequently, $\hat{H}_{J}$ alone determines the spin ground state of $\hat{H}_{\textrm{tJ}}$.
In the classical limit, $\hat{H}_{J}$ is minimized by a 120$^\circ$ spin order such as sketched in Fig.~\ref{fig_effective_hopping}.
Replacing $\hat{\bm{{\cal{S}}}}_{\bm{R}}$ by $\langle\hat{\bm{{\cal{S}}}}_{\bm{R}}\rangle=\frac{\langle\hat{\bm{\sigma}}_{\bm{R}}\rangle}{2}$, this classical order on the $A$, $B$, and $C$ sublattices reads
\begin{equation} \label{eq:Classical_Neel_order_ansatz}
\langle\hat{\bm{\sigma}}_{\bm{R}}\rangle
\equiv
\hat{\bm{n}}_{\bm{R}}
=
\begin{cases}
\bm{e}_x, \; & \bm{R} \in A
\\ 
-\frac{\bm{e}_x}{2}-\frac{\sqrt{3}\bm{e}_y}{2}, \; & \bm{R} \in B
\\
-\frac{\bm{e}_x}{2}+\frac{\sqrt{3}\bm{e}_y}{2}, \; & \bm{R} \in C
\end{cases}.
\end{equation}

To include a low density of spin fluctuations on top of this background order, we rewrite the spinons $\hat{s}_{\bm{R},\tau}$ in terms of Holstein-Primakoff (HP) bosons $\hat{a}_{\bm{R}}$~\cite{Auerbach1994}:
\begin{equation}
\label{eq:spinon_HP_boson_transform}
\hat{U}_{\bm{R}}
\begin{bmatrix} \hat{s}_{\bm{R} \uparrow} \\ \hat{s}_{\bm{R} \downarrow} \end{bmatrix}
=
\begin{bmatrix}
\sqrt{2S-\hat{a}_{\bm{R}}^\dagger\hat{a}_{\bm{R}}}
\\
\hat{a}_{\bm{R}}
\end{bmatrix}.
\end{equation}
Here $\hat{U}_{\bm{R}}$ is the spin rotation matrix from $\bm{e}_z$ to $\hat{\bm{n}}_{\bm{R}}$.
$S$ denotes the spin magnitude.
Although we are ultimately considering $S=\frac{1}{2}$, it is useful to compare with the semi-classical regime $S \gg 1$~\cite{Auerbach1994}.
Magnetic order in this calculation is characterized by the (normalized) sublattice magnetization:
\begin{equation}
\label{eq:sublattice_magnetization}
\begin{aligned}
m
&\equiv
\left[
1-\frac{1}{S}\langle\hat{a}_{\bm{R}}^\dagger\hat{a}_{\bm{R}}\rangle
\right].
\end{aligned}
\end{equation}
We refer to Appendix~\ref{Appendix_Holstein_Primakoff} for further details.

Thus far, all transformations have been exact (except for the perturbation theory used to derive the t-J model).
To make further progress, we consider either of two similar approximations.
First is the standard linear spin-wave (LSW) theory~\cite{Auerbach1994,Martinez1991Spin,Azzouz1996Motion,Chen2022proposal}, namely expanding in $1/S$ and neglecting all subleading terms.
Even though $S = 1/2$ is far from the large-$S$ limit, it has been observed that this approximation still gives the correct qualitative features of spin waves~\cite{Huse1988Simple,Trivedi1989Green}.
Second is a mean-field approximation in which we replace $\sqrt{2S-\hat{a}_{\bm{R}}^\dagger\hat{a}_{\bm{R}}}$ in Eq.~\eqref{eq:spinon_HP_boson_transform} by $\xi \equiv \sqrt{2S-\frac{1}{N}\sum_{\bm{R}}\langle\hat{a}_{\bm{R}}^\dagger\hat{a}_{\bm{R}}\rangle}$, where the expectation value is in the ground state of $\hat{H}_J$.
The value $\langle \hat{a}_{\bm{R}}^{\dag} \hat{a}_{\bm{R}} \rangle$ is then determined self-consistently.

Both approaches ultimately approximate $\hat{H}_J$ by a quadratic Hamiltonian, which a Bogoliubov rotation then diagonalizes.
The resulting expression is, in terms of momenta $\bm{q}$,
\begin{equation}
\label{eq:working_H_J}
\hat{H}_{\textrm{J}}
= \frac{3J \xi^2}{2} \sum_{\bm{q}}
\Omega_{\bm{q}}
\hat{\beta}_{\bm{q}}^\dag\hat{\beta}_{\bm{q}}
, \qquad \hat{\beta}_{\bm{q}} \equiv u_{\bm{q}} \hat{a}_{\bm{q}} - v_{\bm{q}} \hat{a}_{-\bm{q}}^{\dag},
\end{equation}
where
\begin{equation} \label{eq:spinon_frequencies}
\Omega_{\bm{q}} = \sqrt{\left(1+\frac{\gamma_{\bm{q}}}{6}\right)^2-\frac{\gamma_{\bm{q}}^2}{4}},
\end{equation}
\begin{equation} \label{eq:gamma_coefficient}
\gamma_{\bm{q}}
=
\sum_{i=1}^3
\cos(a_M \bm{q}\cdot\bm{e}_i)
,
\end{equation}
\begin{equation} \label{eq:Bogoliubov_u_coefficient}
u_{\bm{q}}
=
\sqrt{\frac{1}{2\Omega_{\bm{q}}}
\left(1+\frac{\gamma_{\bm{q}}}{6}+\Omega_{\bm{q}}\right) },
\end{equation}
\begin{equation} \label{eq:Bogoliubov_v_coefficient}
v_{\bm{q}}
=
\textrm{sgn}[\gamma_{\bm{q}}]
\sqrt{\frac{1}{2\Omega_{\bm{q}}}
\left(1+\frac{\gamma_{\bm{q}}}{6}-\Omega_{\bm{q}}\right) },
\end{equation}
with $\bm{e}_1 = \bm{e}_x$ and $\bm{e}_{2,3} = -\bm{e}_x/2 \pm \sqrt{3} \bm{e}_y/2$ ($\bm{e}_x$ and $\bm{e}_y$ are the $x$ and $y$ unit vectors).
LSW theory corresponds to $\xi^2 = 1$, while the mean-field approximation corresponds to $\xi^2 = (1 + m)S$ (see Eq.~\eqref{eq:sublattice_magnetization}).
In particular, one finds that $m=1-\frac{2}{N}\sum_{\bm{q}}v_{\bm{q}}^2\simeq0.48$ at zero temperature, independent of $t$ and $U$ (see Appendix~\ref{Appendix_Holstein_Primakoff}).

Making the same approximations in $\hat{H}_t$ gives ($N$ denotes the number of moir\'e sites)
\begin{equation} \label{eq:working_H_t}
\begin{aligned}
\hat{H}_t
&=
t\xi^2
\sum_{\bm{k}}
\gamma_{\bm{k}}
\hat{\psi}_{\bm{k}}^\dag
\hat{\psi}_{\bm{k}}
\\
&
+
\frac{\sqrt{3}t\xi}{\sqrt{N}}
\sum_{\bm{k},\bm{q}}
\left[
iM_{\bm{k},\bm{q}}
\hat{\psi}_{\bm{k}+\bm{q}}^\dag
\hat{\psi}_{\bm{k}}
\hat{\beta}_{-\bm{q}}^\dag
+
h.c.
\right]
,
\end{aligned}
\end{equation}
with vertex
\begin{equation} \label{eq:spinon_charge_coupling}
M_{\bm{k},\bm{q}}
=
h_{\bm{k}}v_{\bm{q}}-h_{\bm{k}+\bm{q}}u_{\bm{q}},
\end{equation}
\begin{equation} \label{eq:h_coefficient}
h_{\bm{k}}
\equiv
\sum_{i=1}^3
\sin(a_M \bm{k}\cdot\bm{e}_i).
\end{equation}

Note that the bare holon hopping in Eq.~\eqref{eq:working_H_t} has the opposite sign compared to that of the original hole, which is $-2t\gamma_{\bm{k}}\hat{h}_{\bm{k},\tau}^\dagger\hat{h}_{\bm{k},\tau}$ (Eq.~\eqref{eq:tJ_model_def}).
The minus sign comes from the fact that holon hopping has an additional factor of the dot product between neighboring spin axes [see Eq.~\eqref{eq:slave_particle_H_t}], which is $\cos{2\pi/3} = -1/2$ for 120$^\circ$ order.
Depletion of the magnetization due to spin fluctuations gives a further factor $\xi^2$.


\subsection{Self-consistent Born approximation}

\begin{figure}[t]
\centering
\includegraphics[width=1.0\columnwidth]{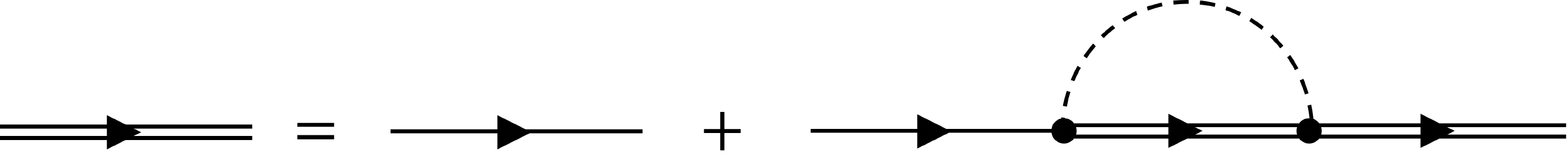}
\caption{Diagrammatic equation defining the self-consistent Born approximation (SCBA) for the holon propagator.
Solid single lines are the bare holon propagator $G_{\bm{k}}^0(\epsilon)$. Solid double lines are the dressed holon propagator $G_{\bm{k}}(\epsilon)$.
The dashed line represents the propagator for a Holstein-Primakoff spin excitation, and black dots indicate the holon-spin vertex (second line of Eq.~\eqref{eq:working_H_t}).}
\label{fig_SCBA}
\end{figure}

\begin{figure}[t]
\centering
\includegraphics[width=0.7\columnwidth]{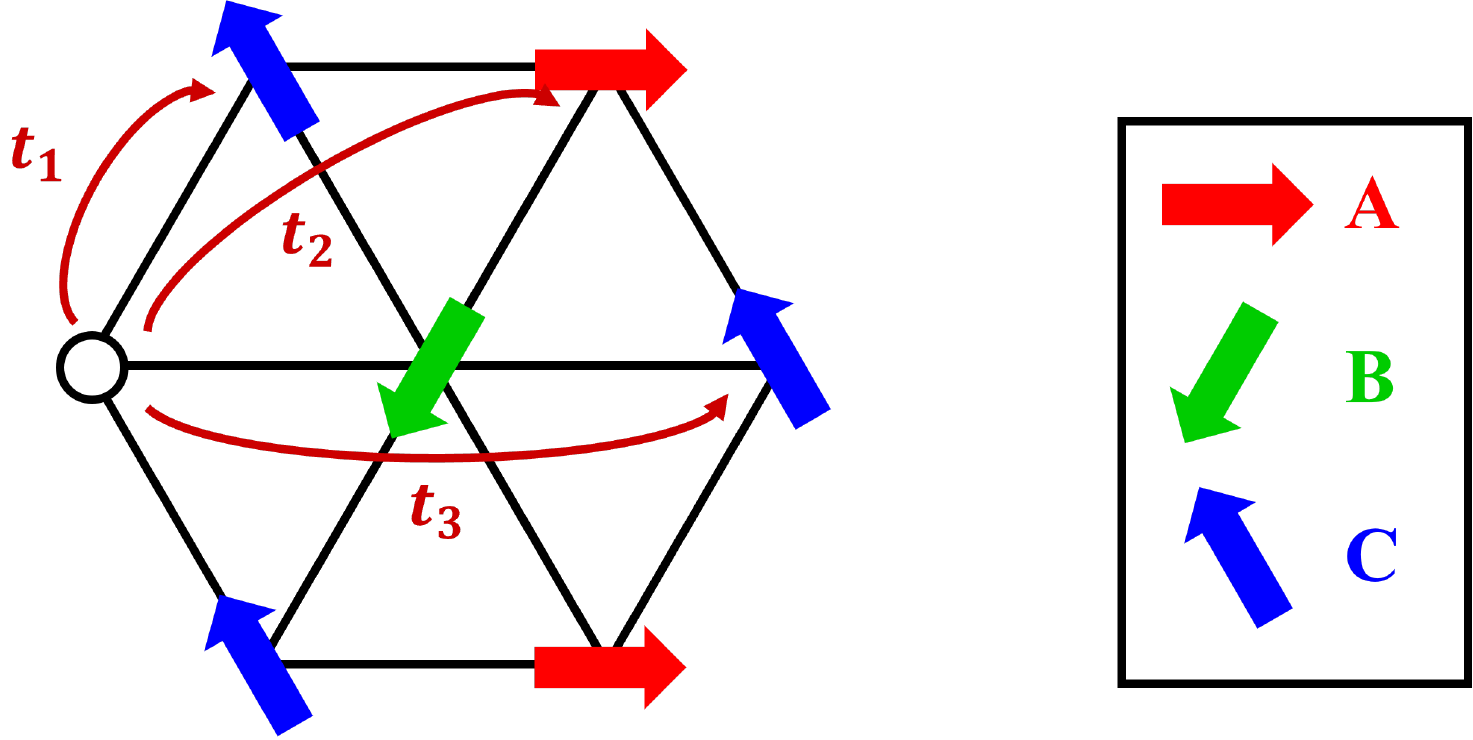}
\caption{Illustration of hopping parameters $t_1$, $t_2$ and $t_3$ for the approximated dressed-holon dispersion in Eq.~\eqref{eq:approximate_single_particle_dispersion}. 
Arrows show the background magnetic order in which the holon (white dot) hops.}
\label{fig_effective_hopping}
\end{figure}

The second term of Eq.~\eqref{eq:working_H_t} leads to a modification of the holon propagator, which we describe via the standard self-consistent Born approximation (SCBA)~\cite{Martinez1991Spin,Azzouz1996Motion,Chen2022proposal} as given in Fig.~\ref{fig_SCBA}.
The SCBA ignores vertex corrections and crossed diagrams, and uses the bare spin propagator corresponding to Eq.~\eqref{eq:working_H_J}.
Fig.~\ref{fig_SCBA} translates to the integral equation
\begin{equation} \label{eq:SCBA_self_energy}
\Sigma_{\bm{k}}(\epsilon)
=
\frac{3t^2 \xi^2}{N} 
\sum_{\bm{q}}
\frac{M_{\bm{k},\bm{q}}^2}{\epsilon-\omega_{\bm{q}}-t\xi^2\gamma_{\bm{k}+\bm{q}}-\Sigma_{\bm{k}+\bm{q}}(\epsilon-\omega_{\bm{q}})},
\end{equation}
where $\Sigma_{\bm{k}}(\epsilon)$ is the dressed holon self-energy.
We solve Eq.~\eqref{eq:SCBA_self_energy} numerically, and determine the effective holon dispersion $\epsilon_{\bm{k}}$ by locating a pole in the propagator (which amounts to solving $\Sigma_{\bm{k}}(\epsilon_{\bm{k}})+t\xi^2\gamma_{\bm{k}}=\epsilon_{\bm{k}}$).

We find that in practice, the effective holon dispersion can be approximated reasonably well by that of the following effective Hamiltonian:
\begin{equation} \label{eq:approximate_single_particle_Hamiltonian}
\begin{aligned}
\hat{H}_{\textrm{d.h.}}
=
-\sum_{\bm{R}}
\sum_{i=1}^3
\bigg[
&
t_1
\hat{\psi}_{\bm{R}}^\dag
\hat{\psi}_{\bm{R}+\bm{e}_i}
+
t_2
\hat{\psi}_{\bm{R}}^\dag
\hat{\psi}_{\bm{R}+\bm{e}'_i}
\\
&\qquad +
t_3
\hat{\psi}_{\bm{R}}^\dag
\hat{\psi}_{\bm{R}+2\bm{e}_i}
\bigg]
+
h.c.
,
\end{aligned}
\end{equation}
where $\bm{e}_i$ are again the nearest-neighbor vectors defined below Eq.~\eqref{eq:Bogoliubov_v_coefficient}, and $\bm{e}'_i$ are next-nearest-neighbor vectors: $\bm{e}'_1=\sqrt{3}\bm{e}_y$ and $\bm{e}'_{2,3}=\pm\frac{3}{2}\bm{e}_x-\frac{\sqrt{3}}{2}\bm{e}_y$.
The dispersion corresponding to Eq.~\eqref{eq:approximate_single_particle_Hamiltonian} is
\begin{equation} \label{eq:approximate_single_particle_dispersion}
\begin{aligned}
\epsilon_{\bm{k}}
&=
-2t_1\gamma_{\bm{k}}
-2t_2\gamma'_{\bm{k}}
-2t_3\gamma_{2\bm{k}},
\end{aligned}
\end{equation}
where $\gamma'_k=\sum_{i=1}^3\cos(a_M \bm{k}\cdot\bm{e}'_i)$.

Eqs.~\eqref{eq:approximate_single_particle_Hamiltonian} and~\eqref{eq:approximate_single_particle_dispersion} have a simple physical interpreation: in addition to the original nearest-neighbor hopping (with renormalized amplitude $t_1$), there is effective hopping to next-nearest-neighbor sites, which can be either to the same or different sublattices (with amplitudes $t_2$ and $t_3$ respectively).
This is illustrated in Fig.~\ref{fig_effective_hopping}.
In our subsequent calculations, we use $\epsilon_{\bm{k}}$ as given by Eq.~\eqref{eq:approximate_single_particle_dispersion} for the holon dispersion, with the hopping amplitudes determined by a fit to the numerical solution of Eq.~\eqref{eq:SCBA_self_energy}.

\subsection{Exciton Hamiltonian}

Recall the effective Hamiltonian given in Eq.~\eqref{eq:two_body_Hamiltonian} of Sec.~\ref{sec:summary} (reproduced here):
\begin{equation} \label{eq:two_body_Hamiltonian_reproduced}
\begin{aligned}
\hat{H}
&=
\sum_{\bm{k}}
\epsilon_{\bm{k}}
\hat{\psi}_{\bm{k}}^\dag
\hat{\psi}_{\bm{k}}
-
2t
\sum_{\bm{k},\tau}
\gamma_{\bm{k}}
\hat{c}_{\bm{k},\tau}^\dag
\hat{c}_{\bm{k},\tau}
\\
&
\quad -
\frac{1}{{\cal{A}}}
\sum_{\tau}
\sum_{\bm{k},\bm{k}',\bm{q}}
V(q)
\hat{c}_{\bm{k}+\bm{q},\tau}^\dag
\hat{\psi}_{\bm{k}'-\bm{q}}^\dag
\hat{\psi}_{\bm{k}'}
\hat{c}_{\bm{k},\tau}
.
\end{aligned}
\end{equation}
The preceding subsections have explained the term $\epsilon_{\bm{k}} \hat{\psi}_{\bm{k}}^{\dag} \hat{\psi}_{\bm{k}}$, and the term $-2t \gamma_{\bm{k}} \hat{c}_{\bm{k}, \tau}^{\dag} \hat{c}_{\bm{k}, \tau}$ is simply the bare CmB hopping term written in momentum space.
To obtain the second line, we take the Coulomb interaction from our starting Hamiltonian --- $V_{|\bm{R} - \bm{R}'|} \hat{c}_{\bm{R}, \tau}^{\dag} \hat{h}_{\bm{R}', \tau'}^{\dag} \hat{h}_{\bm{R}', \tau'} \hat{c}_{\bm{R}, \tau}$ --- and use the constraints on the slave particles (Eq.~\eqref{eq:slave_particle_constraint}) to express $\sum_{\tau'} \hat{h}_{\bm{R}', \tau'}^{\dag} \hat{h}_{\bm{R}', \tau'} = 1 + \hat{\psi}_{\bm{R}'}^{\dag} \hat{\psi}_{\bm{R}'}$.
The constant term amounts to a shift of chemical potential (and should be balanced against the background positive charges in any case), thus we ignore it and are left with Eq.~\eqref{eq:two_body_Hamiltonian_reproduced} in momentum space.

We take the Coulomb interaction to be
\begin{equation}
\label{eq:Coulomb_gate_screening}
V(q)
=
\frac{2\pi e^2}{\epsilon_r}\frac{\tanh{(qd)}}{q}.
\end{equation}
The factor $\tanh{(qd)}$ comes from considering there to be metallic gates at a perpendicular distance $d$ from the TMD bilayer~\cite{Bultinck2020Ground,Chubukov2017Superconductivity}, which screen the charges at distances greater than $d$ (momenta less than $d^{-1}$).
We set $d \gg a_M$, and have found that our results are insensitive to the precise value.
The remaining factors in Eq.~\eqref{eq:Coulomb_gate_screening} are simply the bare interaction for charges forced within a 2D plane.

Although written in second quantization, Eq.~\eqref{eq:two_body_Hamiltonian_reproduced} in the one-electron \& one-holon subspace is a two-body Hamiltonian and can readily be diagonalized numerically.
This gives a set of exciton energies $E_{n, \bm{Q}}^X$ and wavefunctions $\phi_{\bm{Q}}^{(n)}(\bm{p})$, where $\bm{Q}$ and $\bm{p}$ are respectively the total and relative momenta of the electron-holon pair, and $n$ is a discrete index labeling the eigenstates at given $\bm{Q}$ (note that the eigenstates are degenerate with respect to the CmB spin $\tau$).
In particular, the energies and wavefunctions solve the following eigenvalue problem:
\begin{equation}
\label{eq:Wannier_equation}
\sum_{\bm{q}}
\left[
\varepsilon_{\bm{Q}}(\bm{p}) \delta_{\bm{q},0}
-\frac{1}{{\cal{A}}}
V(q)
\right]
\phi_{\bm{Q}}^{(n)}(\bm{p}-\bm{q})
= E_{n,\bm{Q}}^X \phi_{\bm{Q}}^{(n)}(\bm{p})
,
\end{equation}
where $\varepsilon_{\bm{Q}}(\bm{p})$ denotes the two-particle kinetic energy
\begin{equation}
\label{eq:Two_particle_kinetic_energy}
\varepsilon_{\bm{Q}}(\bm{p})
\equiv
\epsilon_{\frac{\bm{Q}}{2}-\bm{p}}
-2t\gamma_{\frac{\bm{Q}}{2}+\bm{p}}
.
\end{equation}
Eq.~\eqref{eq:Wannier_equation} is the standard Wannier equation for excitons~\cite{Haug2004}, albeit with a modified kinetic energy.
Those eigenvalues lying within the band gap correspond to bound states.

To describe normal moir\'e excitons, we again use Eq.~\eqref{eq:two_body_Hamiltonian_reproduced} but with bare holes in place of holons.
Thus the term $\epsilon_{\bm{Q}/2 - \bm{p}}$ in Eq.~\eqref{eq:Two_particle_kinetic_energy} is replaced by $-2t \gamma_{\bm{Q}/2 - \bm{p}}$, and we otherwise solve Eq.~\eqref{eq:Wannier_equation} as before.

With the exciton wavefunctions in hand, we can define composite boson operators $\hat{X}_{n, \tau}(\bm{Q})^{\dag}$ as in Eq.~\eqref{eq:Exciton_operator}, corresponding to creation of an exciton.
Then (in the dilute limit) the Hamiltonian takes the ``quadratic'' form shown in Eq.~\eqref{eq:Exciton_Hamiltonian}.

\subsection{Exciton-light coupling}

As a final step, we investigate the possibility of optically detecting these excitons.
Optical photons couple to the $\bm{q}=0$ component of the inter-band current operator~\cite{Zeng2021The}, which we show in Appendix~\ref{Appendix_slave_fermion_mean_field_current} to be given by
\begin{equation}
\label{eq:Current_operator}
\hat{\bm{j}}^{(cv)}
=
\frac{ev_F}{c}
\sum_{\bm{k},\tau}
\bm{e}_\tau
\hat{c}_{\bm{k},\tau}^\dagger
\hat{h}_{\bm{k},\tau}^\dagger
+
h.c.
,
\end{equation}
with polarization vector $\bm{e}_\tau \equiv \tau \bm{e}_x - i \bm{e}_y$. 
Note that as a result of this polarization vector, circularly-polarized light couples selectively to individual spin/valley $\tau$.
This is true for both moir\'e and Mott-moir\'e excitons~\cite{Rivera2018Interlayer}, since the Hubbard interaction does not enter into the current operator.

To derive further selection rules for Mott-moir\'e excitons, we rewrite Eq.~\eqref{eq:Current_operator} in terms of the exciton operators $\hat{X}_{n, \tau}(\bm{Q})^{\dag}$ (see Appendix~\ref{Appendix_slave_fermion_mean_field_current}).
For definiteness, we consider photons with the polarization vector of $\bm{e}_-$, and the corresponding longitudinal optical conductivity obtained from linear response theory is given by~\cite{Marel2004Optical}, 
\begin{equation}
\label{eq:Optical_conductivity_Mott}
\sigma(\omega)
\sim
\frac{i}{\omega}
\sum_n
\frac{|\Phi_{\bm{\kappa}}^{(n)}|^2}{\omega-E_{n,\bm{\kappa}}^X+i\eta}
,
\end{equation}
where $\Phi_{\bm{\kappa}}^{(n)} \equiv \frac{1}{\sqrt{N}}\sum_{\bm{p}} \phi_{\bm{\kappa}}^{(n)}(\bm{p})$ is the wavefunction amplitude at zero (spatial) separation between electron and holon, specifically at total momentum $\bm{\kappa}$ in the mBZ (see Fig.~\ref{fig_exciton_dispersion_colormap}).
$\eta$ is an infinitesimal positive regulator.
An analogous expression holds for moir\'e excitons (see Appendix~\ref{Appendix_moire_exciton_optical_conductivity}).
Importantly, since $\sigma(\omega)$ is proportional to the probability of zero separation between charges, \textit{only s-wave excitons are optically bright}.

%% file: section_4A_Single_holon_properties.tex
\section{Results} \label{sec:results}

\subsection{Single-holon properties}

\begin{figure}[t]
\centering
\includegraphics[width=1.0\columnwidth]{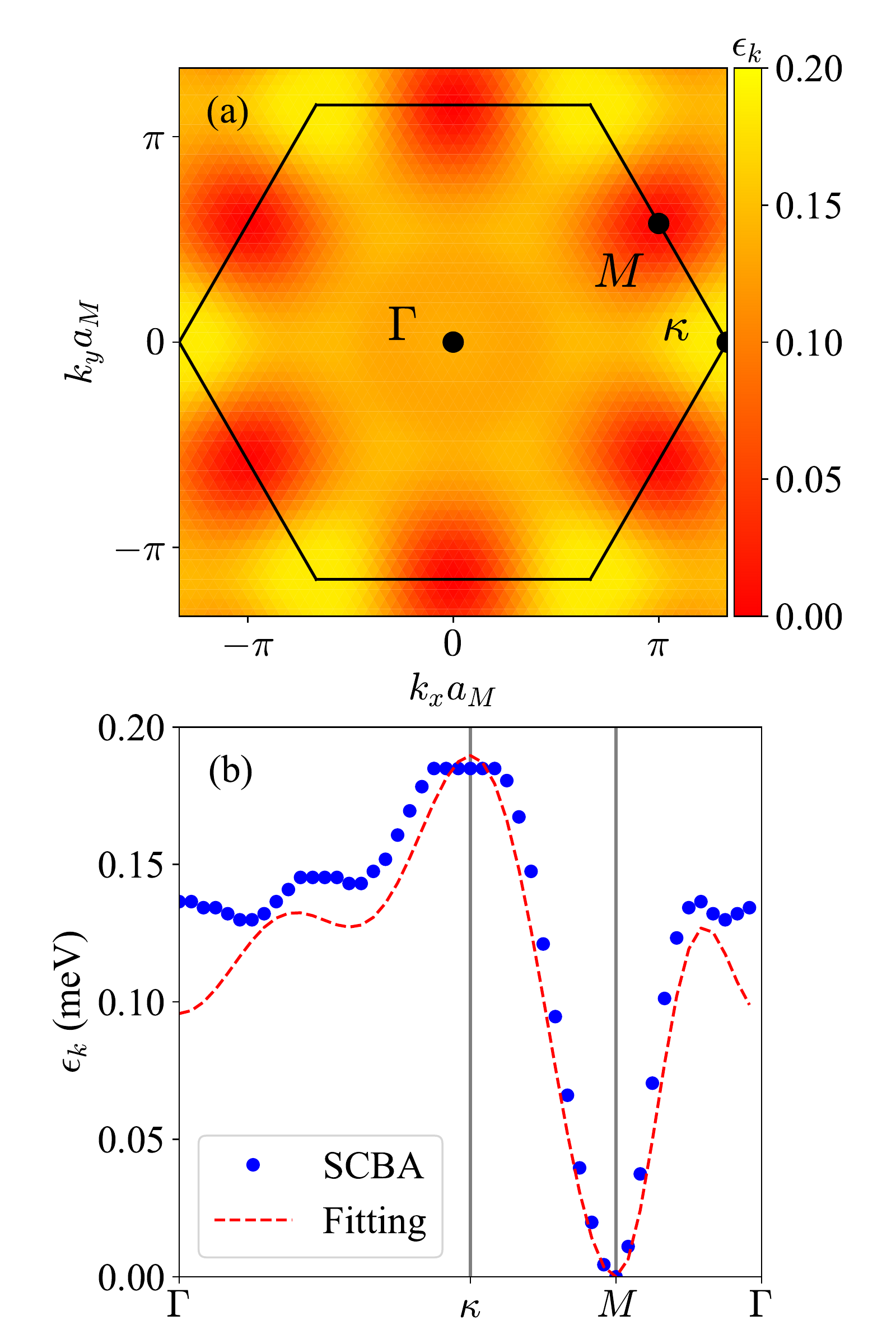}
\caption{Dispersion of dressed holon within the SCBA. 
The moir\'e period is $a_M=10$ nm, at which $t\simeq 1.1$ meV and $J\simeq0.15$ meV according to Ref.~\cite{Wu2018Hubbard}.
Sublattice magnetization is the equilibrium value $m\simeq0.48$. 
System size is $3\times24^2$ sites.
(a) Dispersion $\epsilon_k$ throughout the entire mBZ, indicated by the black hexagon.
Black dots with labels $\Gamma$, $\kappa$, and $M$ indicate important mBZ points.
(b) Line-cut of the dispersion along the path $\Gamma\to\kappa\to M\to\Gamma$.
Blue dots give data from the SCBA, and the red dashed line shows the best fit to Eq.~\eqref{eq:approximate_single_particle_dispersion}.
The minimum of the dispersion is set arbitrarily to zero.}
\label{fig_holon_dispersion}
\end{figure}

We first present results on the properties of individual dressed holons.
Although similar results already exist in the literature~\cite{Azzouz1996Motion,Chen2022proposal}, it is useful to review them here for completeness.

\begin{figure}[t]
\centering
\includegraphics[width=1.0\columnwidth]{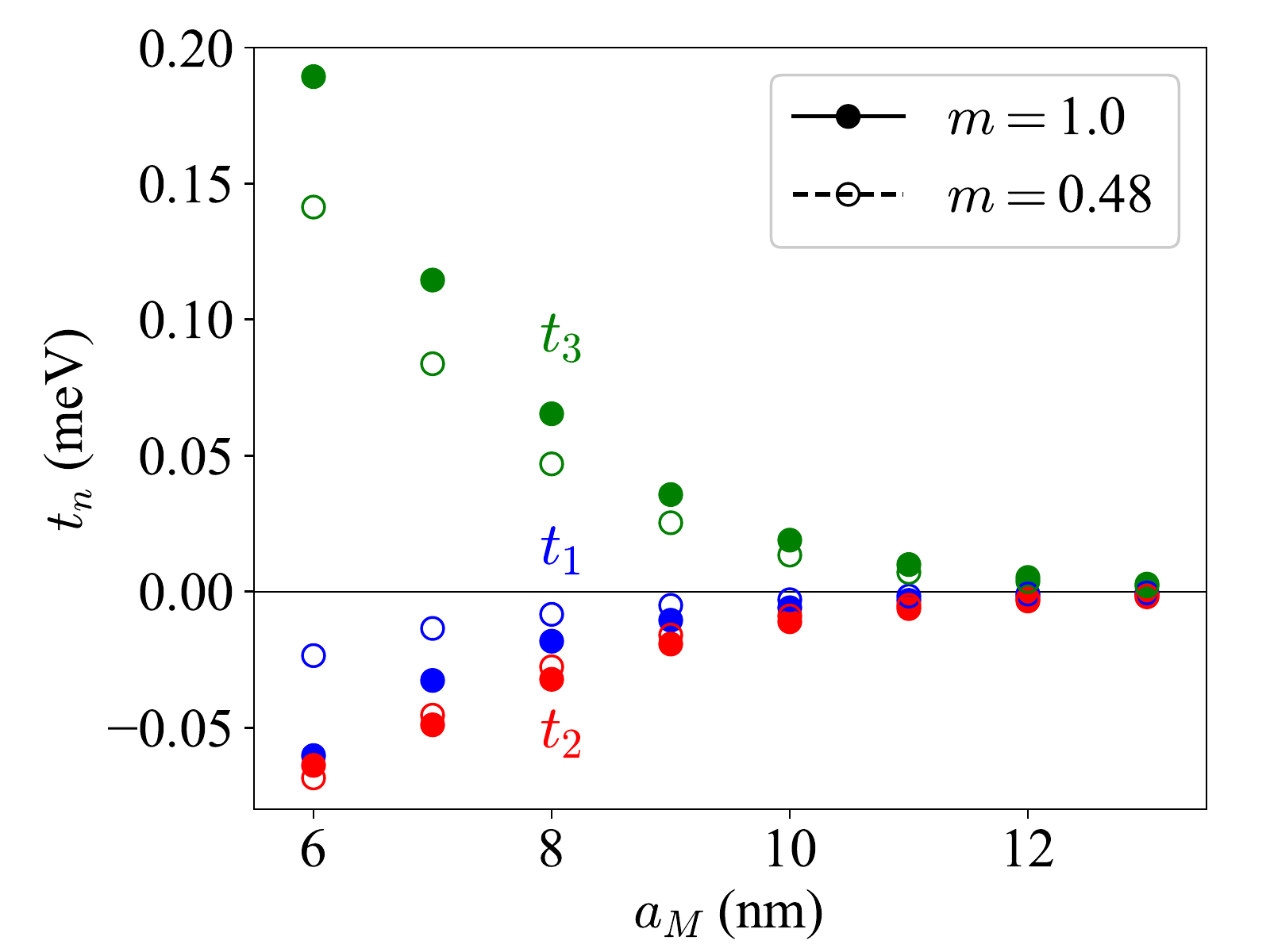}
\caption{Fitting parameters $t_{1,2,3}$ of the dressed holon dispersion in Eq.~\eqref{eq:approximate_single_particle_dispersion} as a function of moir\'e period $a_M$.
System size is $3\times24^2$ sites.
Solid and empty circles represent data for $m=1$ and $m=0.48$ respectively.
Blue, red, and green denote $t_1$, $t_2$, and $t_3$ respectively.
}
\label{fig_holon_hopping_coeff}
\end{figure}

In Fig.~\ref{fig_holon_dispersion}, we show a representative plot of the dressed holon dispersion $\epsilon_{\bm{k}}$ throughout the mBZ.
The minimum is at the point $\bm{M}$ and the maximum is at the point $\bm{\kappa}$, both at the edge of the mBZ.
We also fit to the dispersion of the effective hopping Hamiltonian in Eq.~\eqref{eq:approximate_single_particle_Hamiltonian}, and find reasonable agreement.
The values of the fit parameters $t_{1,2,3}$ as functions of $a_M$ are shown in Fig.~\ref{fig_holon_hopping_coeff}.
Note that, when viewed as functions of $t/J$ (see Appendix~\ref{Appendix_dimensionless}), these results apply to general triangular lattices described by a t-J model and not merely TMD heterobilayers.

The magnitudes of the hopping coefficients decrease significantly as $a_M$ increases. 
The same is true for the holon bandwidth $W$ (see Fig.~\ref{fig_holon_bandwidth}).
We find that $W$ is comparable to $J$, much smaller than the bare hole bandwidth (which scales with $t$).
Qualitatively, this reduction is because a hole in the t-J model is really a magnetic polaron, a charge with a surrounding cloud of spin fluctuations, and the velocity of the polaron is determined by its much slower spin sector~\cite{Grusdt2018Parton}.

Another perspective on the dressed holon dispersion comes from a Hartree-Fock treatment of the triangular-lattice Hubbard model, which we present in Appendix~\ref{Appendix_Hartree_Fock}.
In the large-$U$ limit, the Hartree-Fock Hamiltonian --- which amounts to particles hopping in a Zeeman field determined self-consistently from the average magnetization --- comes out to be precisely of the form in Eq.~\eqref{eq:approximate_single_particle_Hamiltonian}, with parameters $2t_1 = -t + 3J/2$ and $2t_2 = -2t_3 = -3J/4$.
This simple treatment correctly predicts the signs of the effective hoppings, including the extra minus sign in $t_1$, as well as the locations of band extrema obtained from the more sophisticated SCBA.
Interestingly, however, it significantly overestimates the magnitude of the effective hoppings (see Fig.~\ref{fig_holon_hopping_coeff}), and thus we stick to the SCBA results in what follows.


We lastly compare these properties for $m=1$ (LSW theory) to $m=0.48$ (mean-field approximation), also shown in Fig.~\ref{fig_holon_hopping_coeff}.
The magnitudes of $t_{1,2,3}$ are all reduced for the smaller magnetization, but otherwise the behavior is largely unaffected.

%% file: section_4B_Exciton_properties.tex
\subsection{Exciton properties}

We now turn to the properties of Mott-moir\'e excitons obtained by solving Eq.~\eqref{eq:Wannier_equation}.
We compare these results to those of normal moir\'e excitons, where strong correlations in the VmB play no role.
All numerical data uses dielectric constant $\epsilon_r=10$ unless otherwise noted.

\subsubsection{Exciton dispersion}

\begin{figure}[t]
\centering
\includegraphics[width=1.0\columnwidth]{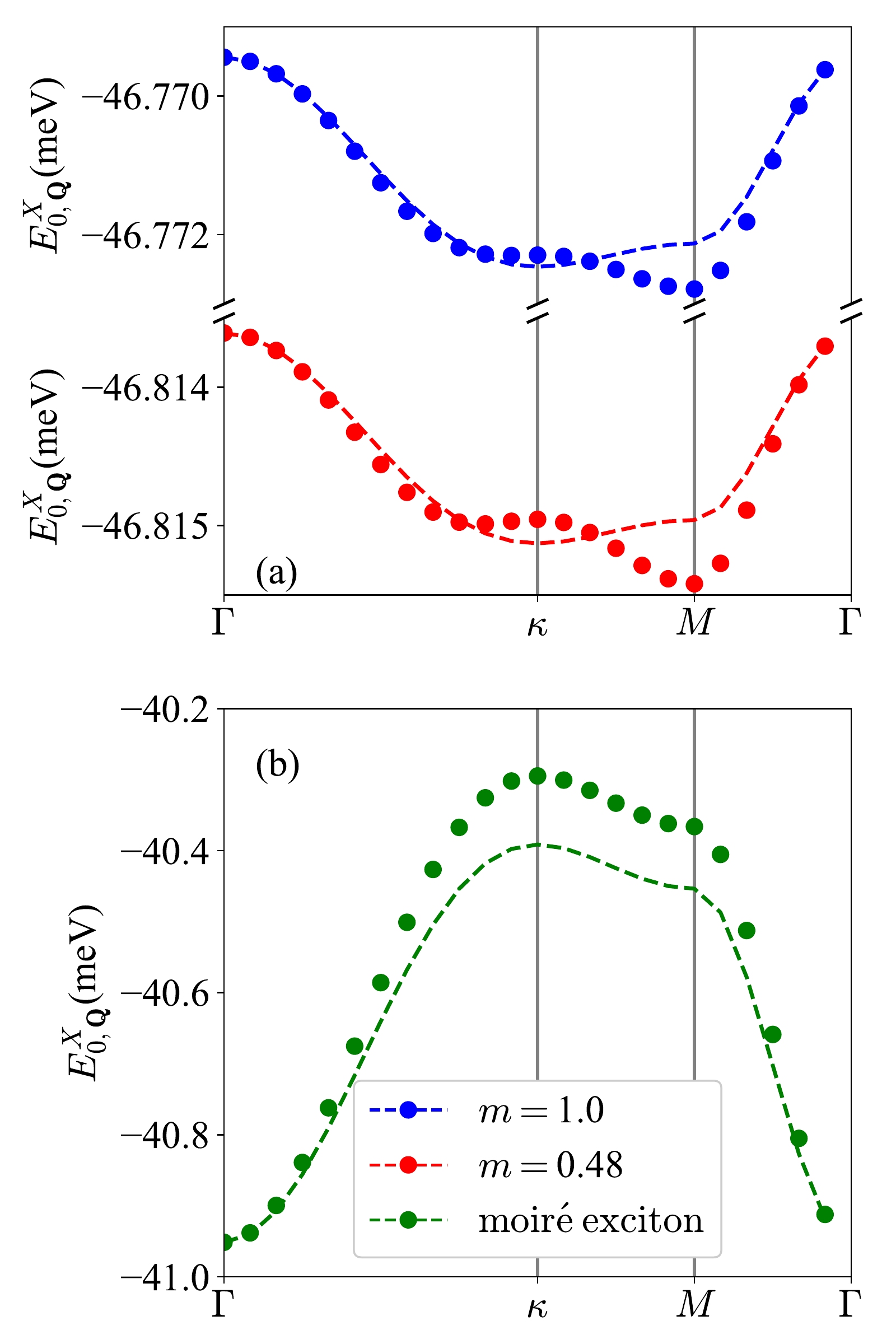}
\caption{Center-of-mass dispersion for the lowest-energy exciton $E_{0,\bm{Q}}^X$, along the path $\Gamma\to\kappa\to M\to\Gamma$. (see Fig.~\ref{fig_exciton_dispersion_colormap}).
Moir\'e period is $a_M = 10$ nm, dielectric constant is $\epsilon_r = 10$, and system size is $3\times 24^2$ sites.
(a) Numerical results for Mott-moir\'e excitons at $m=1$ (blue circles) and $m=0.48$ (red circles).
Dashed lines denote dispersions obtained from perturbation theory, Eq.~\eqref{eq:pertubation_Mott_moire_exciton_dispersion} (shifted so as to coincide with data at the $\Gamma$ point).
(b) Numerical results for moir\'e excitons.
Dashed line again denotes the prediction from perturbation theory, Eq.~\eqref{eq:pertubation_moire_exciton_dispersion}.
}
\label{fig_exciton_dispersion_linecuts}
\end{figure}

First, we discuss the dispersion profile for the lowest-energy moir\'e and Mott-moir\'e excitons (see Figs.~\ref{fig_exciton_dispersion_colormap} and~\ref{fig_exciton_dispersion_linecuts}).
The former has a minimum at $\Gamma$ and a maximum at $\kappa$, whereas the latter has a maximum at $\Gamma$ and a minimum at $M$.
Furthermore, the bandwidth of Mott-moir\'e excitons is drastically narrower than that of moir\'e excitons.

We can understand these differences by noting that at large super-lattice period $a_M$, since the kinetic energy scale decreases exponentially with $a_M$ but the interaction scale decreases only as $1/a_M$, the term $\varepsilon_{\bm{Q}}(\bm{p})$ in Eq.~\eqref{eq:Wannier_equation} can be treated as a perturbation compared to $V(q)$.
At zeroth order, the exciton eigenstates are simply (relative) position eigenstates, since these diagonalize the Coulomb interaction.
Denote the unperturbed eigenstates by $|j, \bm{Q}\rangle$, with $j$ an integer labeling positions in order of increasing separation, and denote the unperturbed energies by $-V_j$.
We give details of the perturbation theory in Appendix~\ref{Appendix_perturbation}, ultimately finding that the first momentum-dependent correction to the moir\'e exciton is $-\frac{t^2}{V_0 - V_1} \gamma_{\bm{Q}}$ whereas that for the Mott-moir\'e exciton is $\frac{t|t_1|}{V_0 - V_1} \gamma_{\bm{Q}}$ (the factor of $t$ comes from the electron hopping and the factor of $t_1$ from the holon).
Note first of all the relative minus sign between the two, and second that the Mott-moir\'e dispersion is reduced by an overall factor of $|t_1|/t$.
Thus we see that both the inverted dispersion and smaller bandwidth of Mott-moir\'e excitons can be traced back to the renormalization of the holon hopping.

Recall that the extra minus sign in $t_1$ --- and thus the inverted Mott-moir\'e dispersion --- can be understood through Hartree-Fock theory, which treats the background spin order as static.
Spin \textit{fluctuations} can therefore be seen as not essential to this phenomenon.
However, they play a much more significant role in the reduced Mott-moir\'e bandwidth, since Hartree-Fock theory alone overestimates the magnitude of $t_1$ --- and thus the bandwidth --- significantly as compared to the SCBA.

\subsubsection{Properties of the lowest exciton state}

\begin{figure}[t]
\centering
\includegraphics[width=1.0\columnwidth]{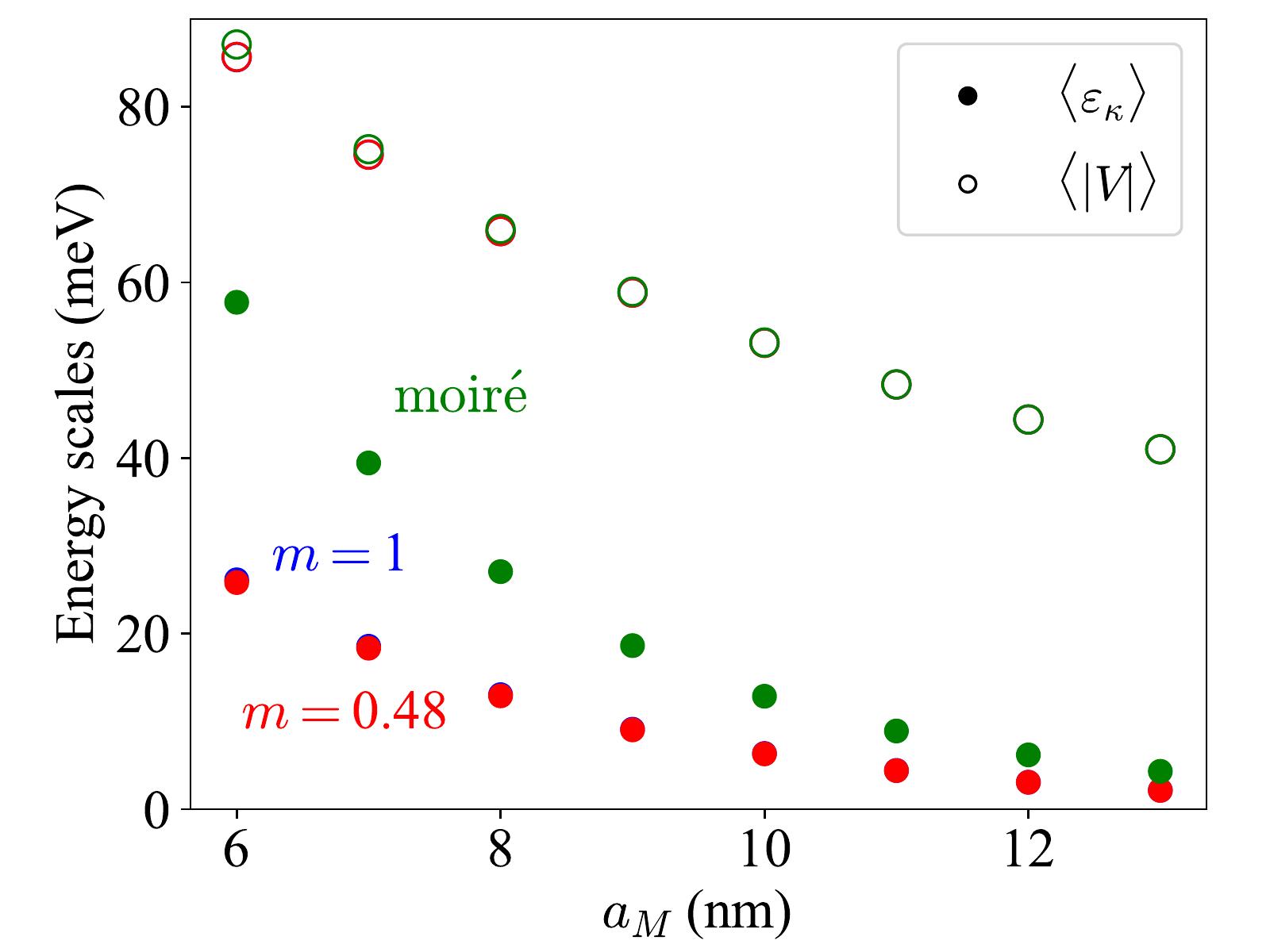}
\caption{Average kinetic and potential energies for the lowest-energy exciton as a function of moir\'e period $a_M$.
Total momentum is set to $\bm{Q} =  \kappa$.
Dielectric constant is $\epsilon_r = 10$, and system size is $3\times 24^2$ sites. 
Blue and red circles are data for Mott-moir\'e excitons at $m=1$ and $m=0.48$ respectively (indistinguishable at this scale).
Green circles are for moir\'e excitons.
Solid markers indicate the two particle kinetic energy $\langle \varepsilon_\kappa \rangle$ (Eq.~\eqref{eq:Two_particle_kinetic_energy}), and empty markers indicate the Coulomb energy ${\cal{A}}^{-1} \langle |V(q)| \rangle$ (Eq.~\eqref{eq:Coulomb_gate_screening} with ${\cal{A}}$ the system area).
}
\label{fig_kinetic_potential}
\end{figure}

\begin{figure}[t]
\centering
\includegraphics[width=1.0\columnwidth]{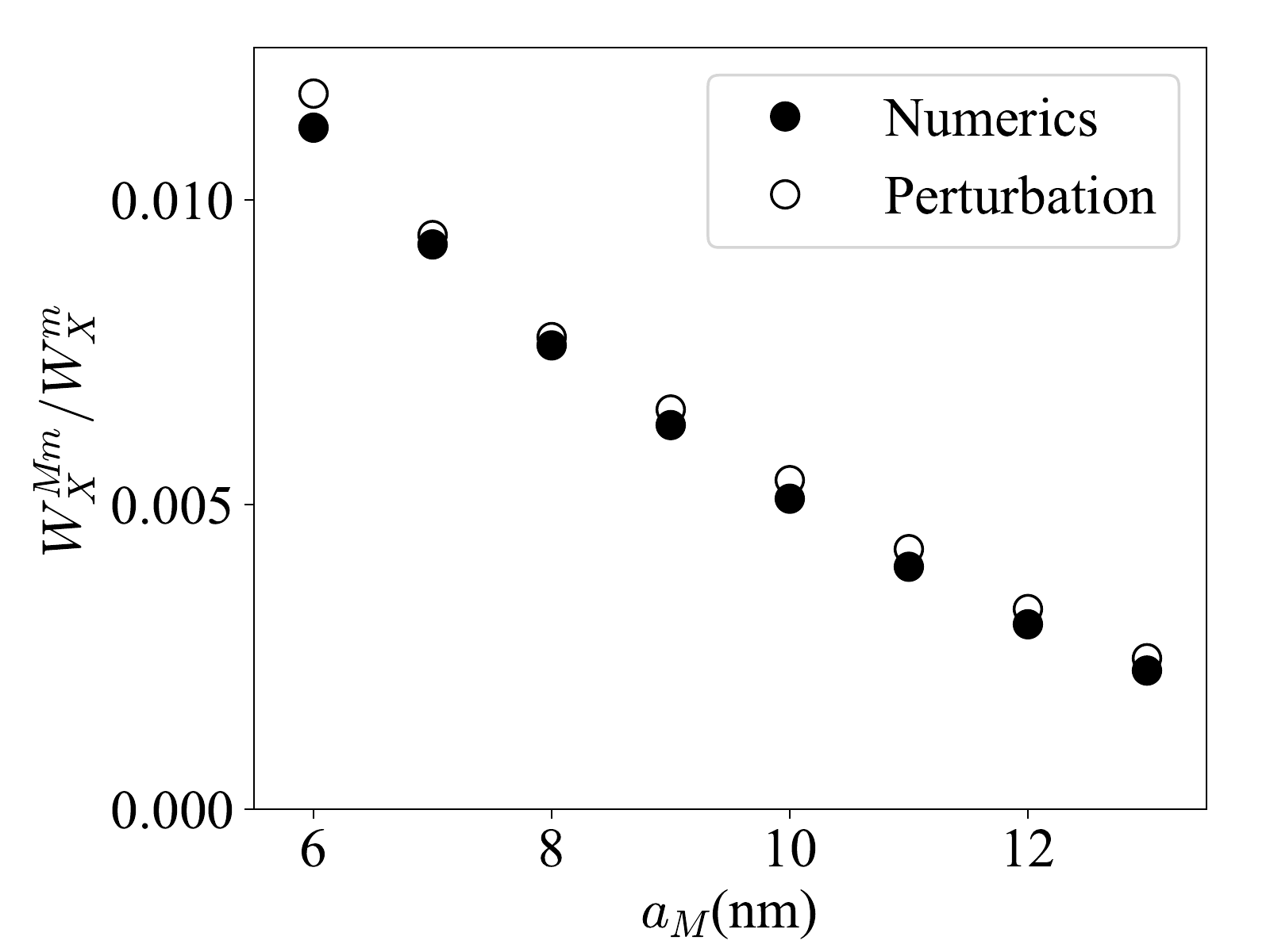}
\caption{Ratio between the bandwidths of Mott-moir\'e ($W_X^{Mm}$) and moir\'e ($W_X^{m}$) excitons from data (solid circles, see also Fig.~\ref{fig_exciton_properties}) and from Eq.~\eqref{eq:bandwidth_ratio_perturbation} (empty circles).}
\label{fig_bandwidth_perturbation}
\end{figure}

We now turn to detailed properties of the lowest-energy excitons, particularly their binding energies, sizes, and bandwidths. 
The results are summarized in Fig.~\ref{fig_exciton_properties}.

Fig.~\ref{fig_kinetic_potential} compares the separate kinetic and potential energies of both excitons.
As argued above, the potential energy is noticeably larger than the kinetic energy within the range of $a_M$ we consider, especially for Mott-moir\'e excitons.
This both explains the small exciton diameters $\langle r\rangle_X\ll a_M$ [see Fig.~\ref{fig_exciton_properties}(b)] and justifies our perturbative treatment outlined in Appendix~\ref{Appendix_perturbation}. 

Since the Coulomb attraction conserves the the total momentum $\bm{Q}$, we define the binding energy as $E_{0,\bm{Q}}^B\equiv \min_{\bm{p}}\varepsilon_{\bm{Q}}(\bm{p})-E_{0,\bm{Q}}^X$, i.e., the difference between $E_{0,\bm{Q}}^X$ and the lowest non-interacting two-particle state at momentum $\bm{Q}$.
The perturbative analysis described above gives $E_{0,\bm{Q}}^B \sim V_0-6t-2t\gamma_{\bm{Q}}$ for moir\'e excitons and $E_{0,\bm{Q}}^B \sim V_0-6t$ for Mott-moir\'e excitons.
This explains the slightly larger binding energy for moir\'e excitons at $\bm{Q}=\bm{\kappa}$ [see Fig.~\ref{fig_exciton_properties}(a)].
We refer to Appendix~\ref{Appendix_perturbation} for more details.

Finally, we elaborate on the bandwidth $W_X$ for both excitons.
We have already discussed how the significantly smaller Mott-moir\'e bandwidth is a consequence of the dressed holon dispersion, but our perturbative analysis makes a further quantitative prediction: the reduction of the bandwidth is $|t_1/t|$ to leading order.
We show in Fig.~\ref{fig_bandwidth_perturbation} that this result is borne out quite well in the numerics.
A further observation is that even the moir\'e bandwidth itself is much smaller than the hopping coefficient $t$ which one might naively expect.
This effect is due to strong Coulomb binding on the lattice~\cite{Mattis1986The}, 
with physical origin given above.


\subsubsection{Excited states and optical spectrum}


\begin{figure}[t]
\centering
\includegraphics[width=1.0\columnwidth]{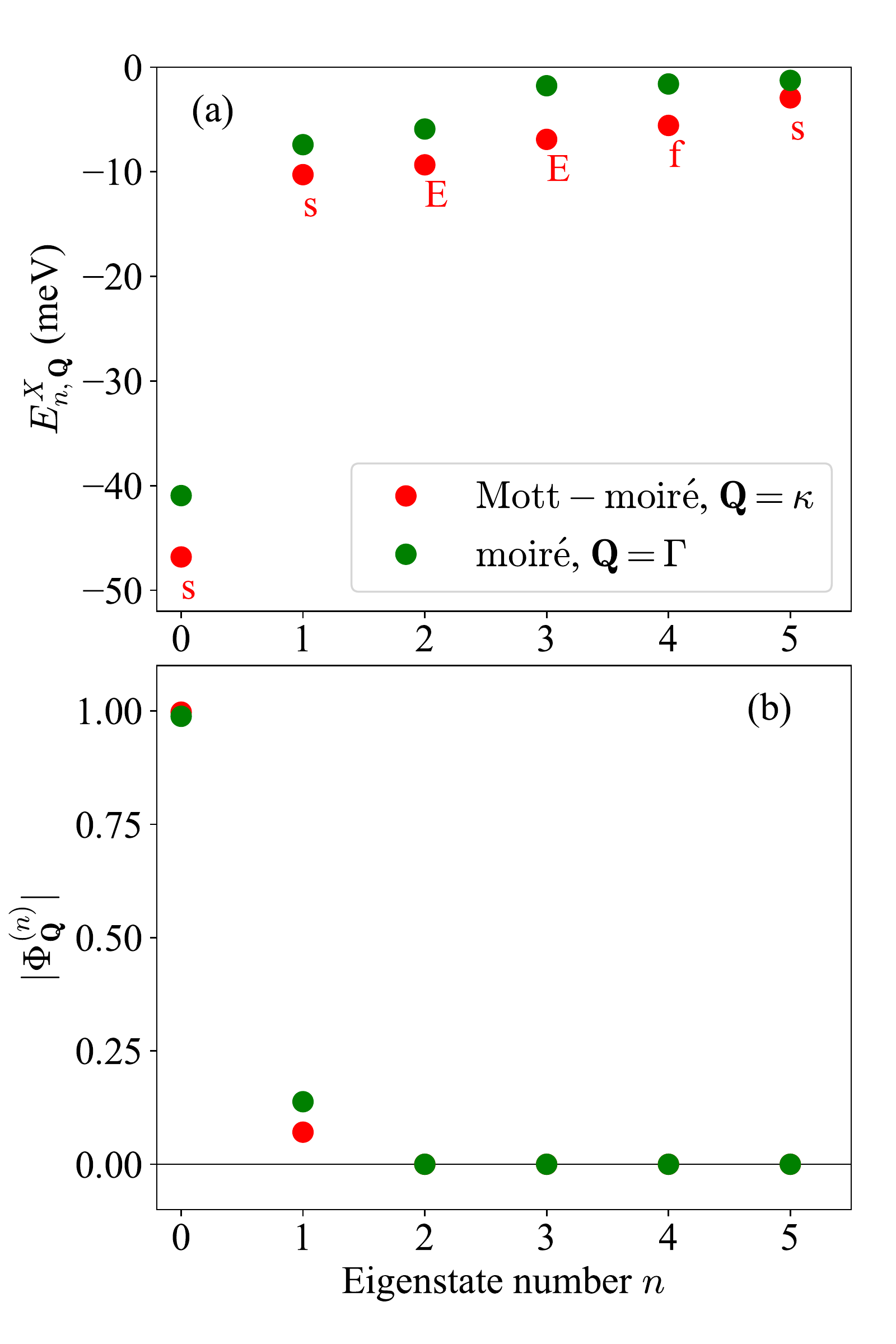}
\caption{(a) Energies $E_{n,\bm{Q}}^X$ of the first few excited states for moir\'e and Mott-moir\'e excitons, at the indicated total momenta. 
The labels $s$, $E$, and $f$ denote the D3 group representations with which the states are associated.
(b) Wavefunction amplitude $|\Phi_{\bm{Q}}^{(n)}|$ which determines the oscillator strength, for the first few excited states of both excitons.
The eigenstate number $n$ simply labels the states (only one of each degenerate pair is shown).
The total momenta used ($\bm{Q}=\bm{\kappa}$ for Mott-moir\'e and $\bm{Q}=\bm{\Gamma}$ for moir\'e) are those which are relevant for the optical conductivity (see Eqs.~\eqref{eq:Optical_conductivity_Mott} and~\eqref{eq:Optical_conductivity_moire}).
Data is for moir\'e period $a_M=10$ nm, dielectric constant $\epsilon_r = 10$, magnetization $m=0.48$, and system size $3 \times 24^2$ sites.
}
\label{fig_exciton_level}
\end{figure}

The first few excited-state exciton energies $E_{n,\bm{Q}}^X$ are shown in Fig.~\ref{fig_exciton_level}(a), at the values of $\bm{Q}$ relevant for the optical conductivity in both cases ($\Gamma$ for moir\'e and $\kappa$ for Mott-moir\'e). 
These levels are not well-described by the Rydberg series $E_{n,\bm{Q}}^X\sim (2n+1)^{-1}$ found in hydrogenic excitons~\cite{Haug2004}, but this is merely a consequence of the lattice structure together with the small exciton radii.
Also note that our use of a two-band model restricts us to excitonic states formed from the valence and conduction moir\'e bands, whereas experimental optical spectra would include contributions from composite particles having constituents in other moir\'e bands.

Fig.~\ref{fig_exciton_level}(b) plots the wavefunction amplitude $\Phi_{\bm{Q}}^{(n)}$ which determines the oscillator strength.
The lowest-energy states substantially dominate the spectra (for both excitons), and furthermore, as discussed in Sec.~\ref{sec:formalism}, only s-wave excitons exhibit a response.


\subsubsection{Exciton wavefunction}

\begin{figure}[t]
\centering
\includegraphics[width=1.0\columnwidth]{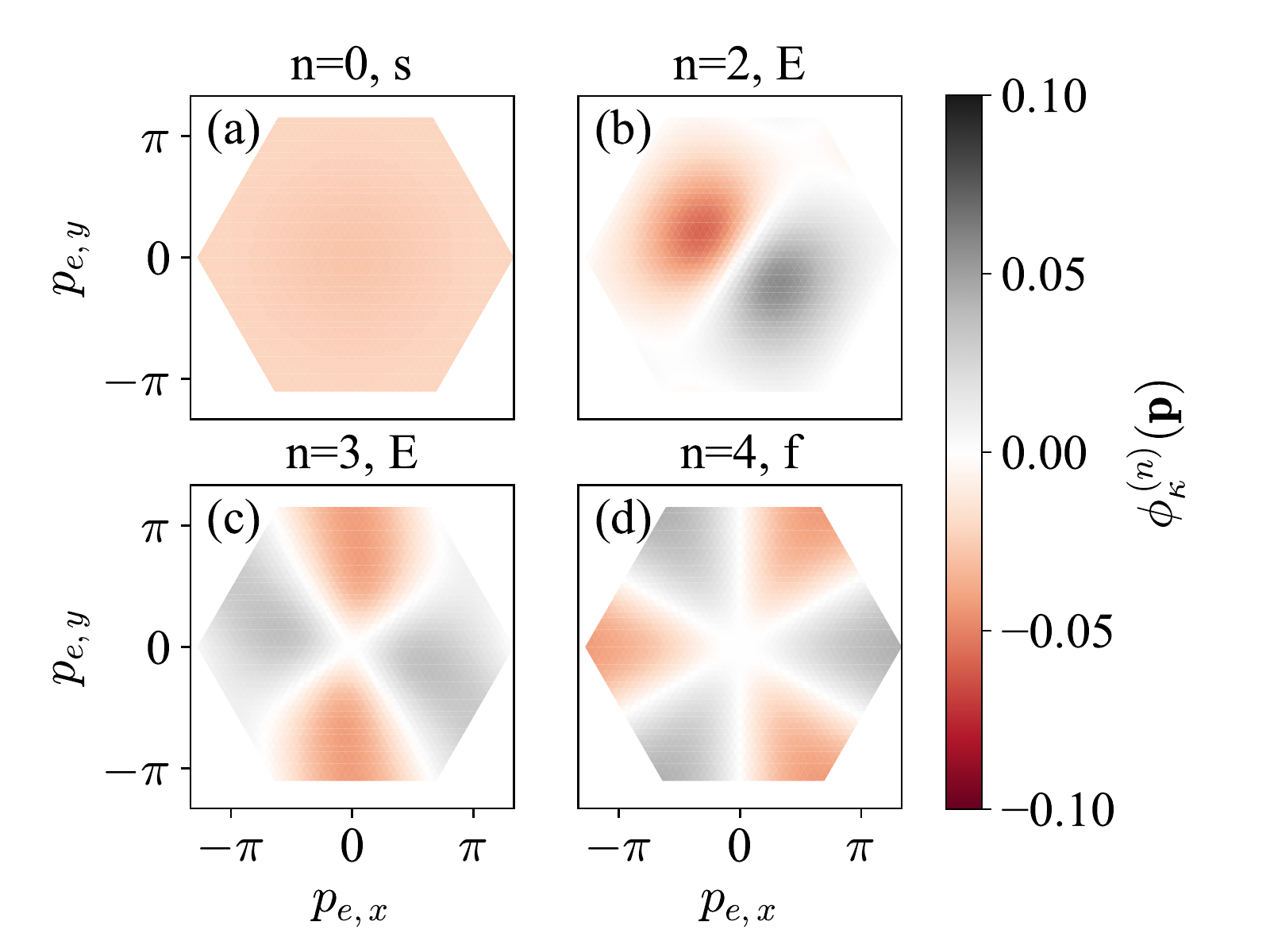}
\caption{The wavefunctions for the first few excited states of Mott-moir\'e excitons.
We set the total momenta $\bm{Q}=\bm{\kappa}$, the moir\'e period $a_M=10$ nm, and the magnetization $m=0.48$ for this figure.
The axes $(p_{e,x},p_{e,y})$ denote the CmB electron momentum of the exciton, within the first mBZ.
The colorbar gives the exciton wavefunction $\phi_\kappa^{(n)}(\bm{p})$.
$n$ in the titles denote the eigenstate number in Fig.~\ref{fig_exciton_level} to which the wavefunctions belong.
The $n=0$ and $n=1$ states are s-waves with different energies.
The $n=1$ state is similar to $n=0$ and is not plotted here.
The p-wave-like state $n=2$ and d-wave-like state $n=3$ are doubly degenerate.
The $n=4$ state is f-wave.
The $s$, $E$, and $f$ labels in the titles denote the D3 group representations with which the states are associated.
}
\label{fig_exciton_WF}
\end{figure}

We show the lowest state wavefunctions of different angular momentum (s-, p-, d- or f-symmetry) for Mott-moir\'e exciton at $\bm{Q}=\bm{\kappa}$ in Fig.~\ref{fig_exciton_WF}.
We understand their rotational properties with the D3 point group symmetry of Eq.~\eqref{eq:Wannier_equation} (in terms of the electron momentum $\bm{p}_e=\frac{\bm{\kappa}}{2}-\bm{p}$).
D3 point group should give two 1-dimensional representations and one 2-dimensional representation~\cite{Tinkham2003Group}.
The 1-dimensional representations can be identified as s- and f-wave states (see Fig.~\ref{fig_exciton_WF}).
The 2-dimensional representation, which we label $E$, cannot be interpreted cleanly in terms of the usual angular momentum classification (Fig.~\ref{fig_exciton_WF} shows examples of apparently p- and d-wave states which both belong to $E$).

%% file: section_5_Conclusion.tex
\section{Conclusion} \label{sec:conclusion}

We have demonstrated the existence of bound states between spin-dressed holons, i.e., magnetic polarons, and conduction electrons on the moir\'e super-lattice of twisted TMD heterobilayers.
Such bound states, named Mott-moir\'e excitons, possess much narrower bandwidths than moir\'e excitons.
Thus the degree of correlations, controllable by gate voltages~\cite{Bultinck2020Ground,Chubukov2017Superconductivity}, offers a further mechanism to engineer exciton properties.
This is in addition to the already high tunability provided by the moir\'e period.
However, we predict that only s-wave excitons (both moir\'e and Mott-moir\'e) are detectable via optical measurements.

These results are a consequence of two simple physical features.
First, the kinetic energy of the holon is heavily suppressed by spin fluctuations.
We further draw a distinction between effects which are due to the presence of \textit{static} spin order versus genuine \textit{fluctuations} --- inversion of the Mott-moir\'e dispersion can be traced to the former, but the reduction of the bandwidth is due to the latter.
Second, the Coulomb energy is much larger than kinetic at large moir\'e periods due to the exponential suppression of the latter.
This allows us to treat the hopping terms as a perturbation, and the exciton properties follow straightforwardly.

One natural question is how to distinguish between moir\'e and Mott-moir\'e excitons experimentally.
Since the main difference is in their masses (i.e., bandwidths), we propose diffusion measurements as one viable possibility.
Intuitively, excitons with larger mass should have slower diffusion, and so diffusion constants should be significantly reduced in the presence of Mott physics.
Recent diffusion measurements have been performed on excitons in TMD heterobilayers~\cite{Choi2020Moire,Li2020Exciton}, but have not compared different VmB fillings (and hence degree of correlations) to the best of our knowledge.

The existing experiments on excitons in the Mott-insulating phase of moir\'e TMDs have focused thus far on effects due to charge order~\cite{Xu2020Correlated,Tang2020Simulation,Regan2020Mott,Campbell2022Strongly}, but not spin order.
Since the energy scale for charge order is $U$ and that for spin order is only $J$, it should be possible to separate these effects by varying the temperature $T$.
Changes that occur at $T \sim J$ can likely be attributed to spin order alone.
We expect that our work, having given a systematic study of the role of spin order on Mott-moir\'e excitons, can inform these future experiments.

Much work remains to be done beyond the inter-band Mott-moir\'e excitons considered here.
For example, at half-filling of the VmB there should also exist \textit{intra-band} excitons consisting of two magnetic polarons [see Fig.~\ref{fig_Mott_moire_exciton_illustrate}(b)].
Previous work has discussed these excitons for a single-band Hubbard model on the square lattice~\cite{Huang2020Spin}, but no such work for triangular moir\'e super-lattices has been done to the best of our knowledge.
The optical properties of strongly-correlated excitons, both inter- and intra-band, such as their coupling to optical cavities and potential cavity-QED effects, are highly active topics as well~\cite{bloch2022strongly,Camacho-Guardian2022Moire}.
Moreover, twisted TMD bilayers show a remarkably rich strong-correlation phase diagram as a function of VmB filling, with both charge and spin order emerging at fractional fillings~\cite{Pan2020quantum}.
The question of excitons in these phases remains open.

%% file: section_6_Acknowledgements.tex
\begin{acknowledgments}
T.-S. H. thanks Daniel Suarez Forero, Supratik Sarkar, and Beini Gao for beneficial experimental discussions. Y.-Z.C. thanks Zhentao Wang for discussions related to slave particles. M.H. thanks Atac Imamoglu and Ajit Srivastava for extensive fruitful discussions. The work at Maryland was supported by JQI-NSF-PFC and Laboratory for Physical Sciences, ARO W911NF2010232, AFOSR FA9550- 19-1-0399 and FA95502010223, and Minta Martin and Simons Foundation. F. W. is supported by National Key Research and Development Program of China 2021YFA1401300 and start-up funding of Wuhan University.
\end{acknowledgments}

%% file: section_A_dimensionless.tex
\section{Results in dimensionless variables} 
\label{Appendix_dimensionless}

\begin{figure}[t]
\centering
\includegraphics[width=1.0\columnwidth]{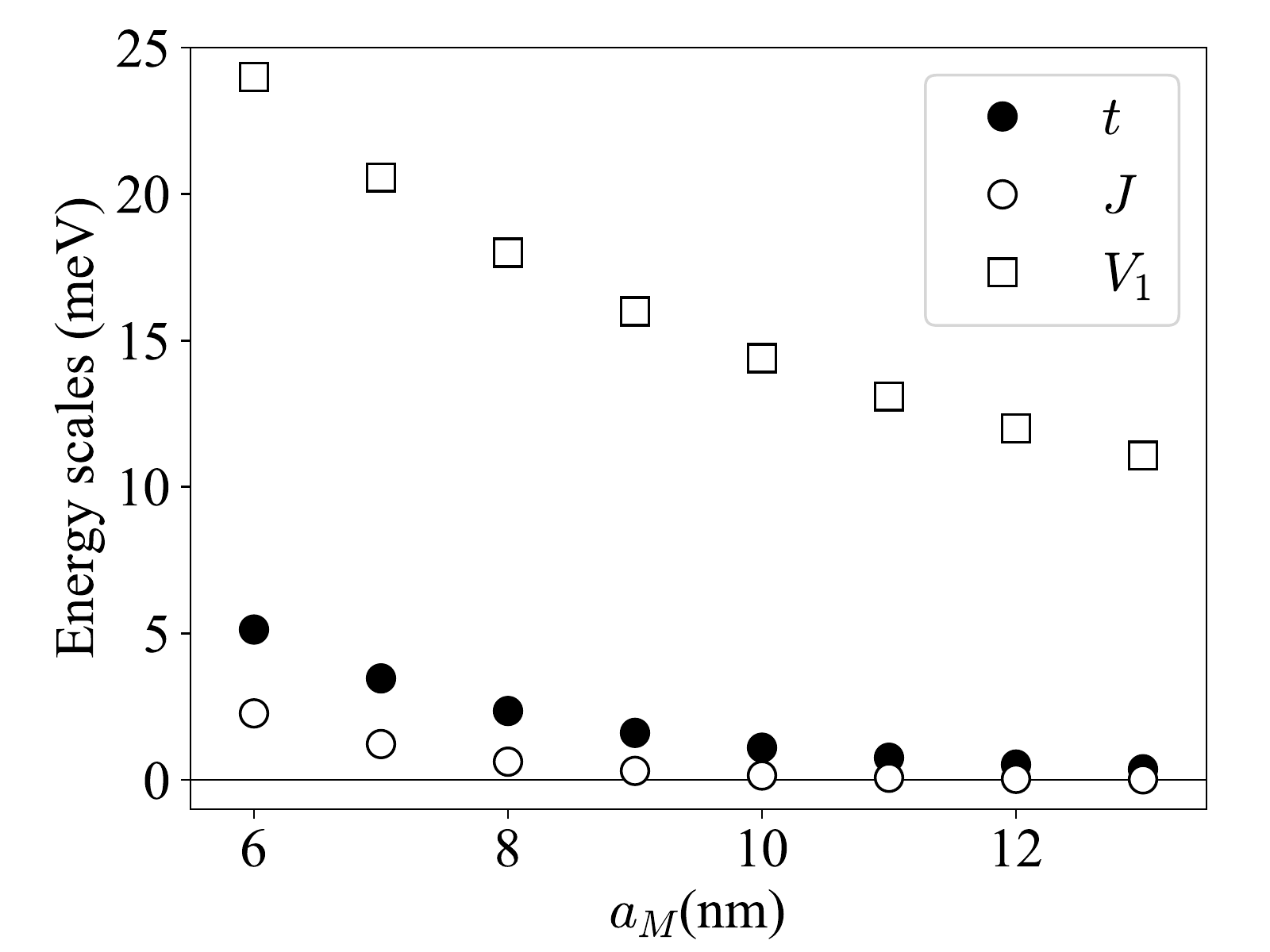}
\caption{Energy scales $t$ (solid circles) and $J$ (empty circles) for the t-J model in Eq.~\eqref{eq:tJ_model_def} as a function of moir\'e period $a_M$, taken from Ref.~\cite{Wu2018Hubbard} for WSe\textsubscript{2} on top of MoSe\textsubscript{2}.
Also shown is the nearest-neighbor Coulomb scale $V_1=\frac{e^2}{\epsilon_ra_M}$ (empty squares), using dielectric constant $\epsilon_r=10$. 
}
\label{fig_Jandt}
\end{figure}

\begin{figure}[t]
\centering
\includegraphics[width=1.0\columnwidth]{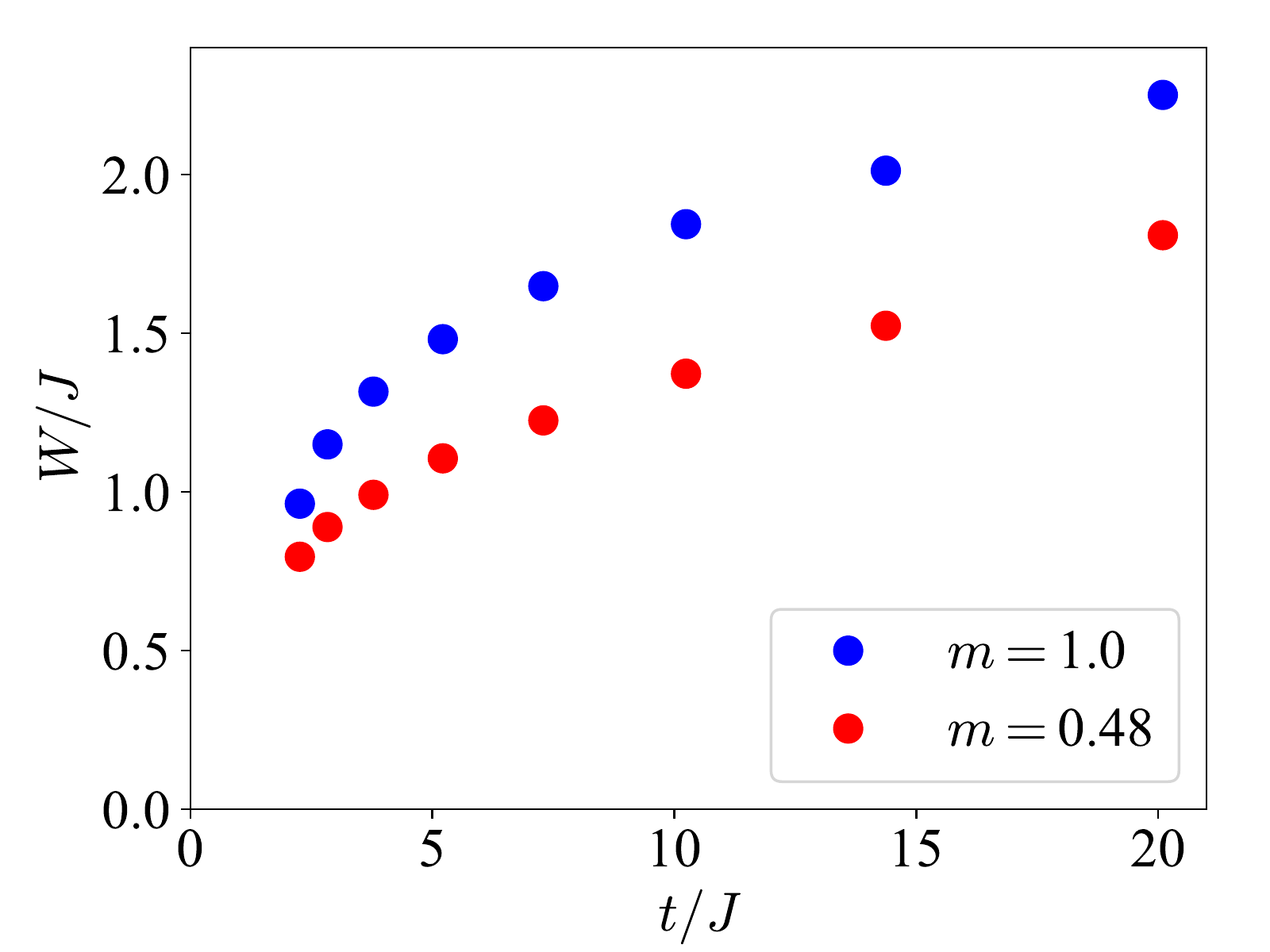}
\caption{Dressed holon bandwidth $W$ (in units of $J$) from SCBA as a function of $t/J$ (values taken from Fig.~\ref{fig_Jandt}), at different sublattice magnetizations $m$ (blue and red).
System size is $3\times24^2$ sites.
}
\label{fig_holon_bandwidth_dimensionless}
\end{figure}

\begin{figure}[t]
\centering
\includegraphics[width=1.0\columnwidth]{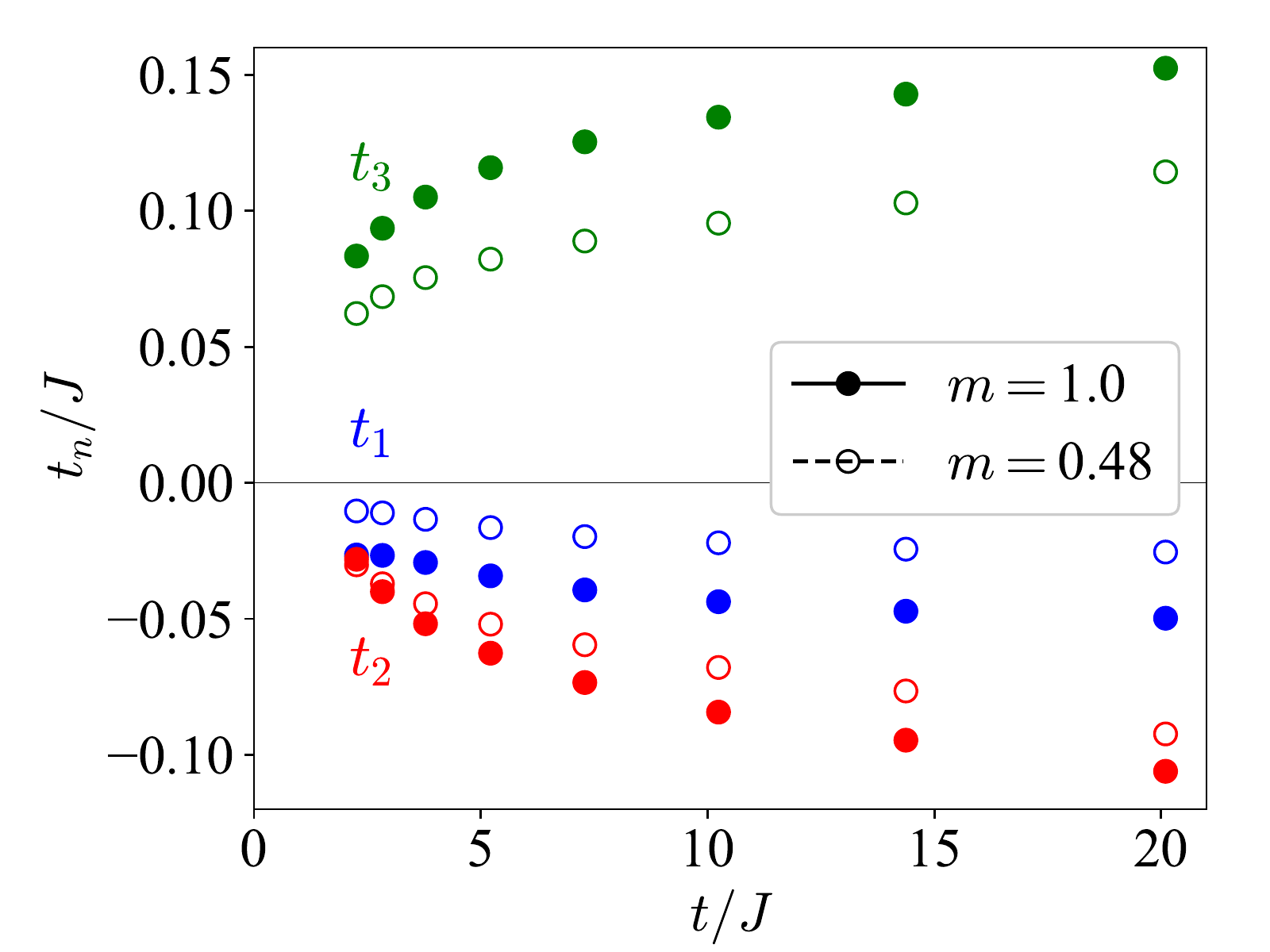}
\caption{Fitting parameters $t_{1,2,3}$ of the dressed holon dispersion (in units of $J$) in Eq.~\eqref{eq:approximate_single_particle_dispersion} as a function of $t/J$ (values taken from Fig.~\ref{fig_Jandt}).
System size is $3\times24^2$ sites.
Solid and empty circles represent data for $m=1$ and $m=0.48$ respectively.
Blue, red, and green denote $t_1$, $t_2$, and $t_3$ respectively.
}
\label{fig_holon_hopping_coeff_dimensionless}
\end{figure}

\begin{figure}[t]
\centering
\includegraphics[width=1.0\columnwidth]{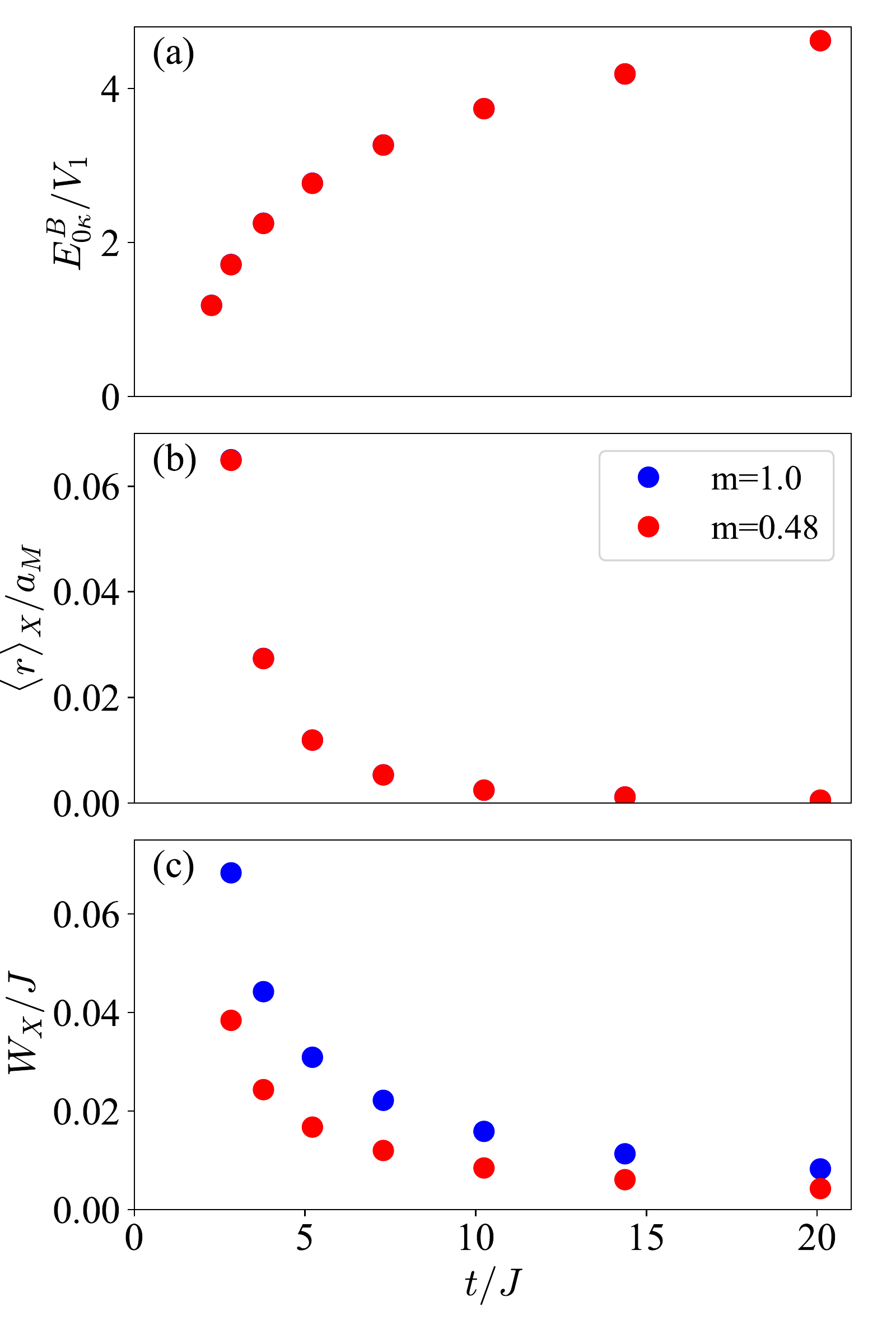}
\caption{Properties of Mott-moir\'e excitons at different magnetization (blue and red, indistinguishable at the scales of top and middle panels) as functions of the moir\'e period $a_M$.
Dielectric constant is $\epsilon_r=10$.
System size is $N=3\times24^2$ sites.
(a) Binding energy of the lowest internal state $E_{0,\kappa}^B$ at total momentum $\bm{Q}=\bm{\kappa}$, which has the largest binding among all $\bm{Q}$.
Energies are in units of $V_1 \equiv \frac{e^2}{\epsilon_ra_M}$, where we fix $a_M=10$nm while varying $t/J$.
(b) Average diameter of excitons at total momentum $\kappa$, in units of $a_M$. 
(c) Exciton bandwidths $W_X$, in units of $J$. 
Values for $t$ and $J$ as functions of $a_M$ are taken from Ref.~\cite{Wu2018Hubbard} for WSe\textsubscript{2} on top of MoSe\textsubscript{2} (see also Fig.~\ref{fig_Jandt}). 
}
\label{fig_exciton_properties_dimensionless}
\end{figure}

Our results apply to generic triangular lattices beyond moir\'e TMDs to which the two band model Eq.~\eqref{eq:two_body_Hamiltonian} applies.
$t$ and $U$ (or $J$) are sufficient to set the dressed holon properties in VmB.
For moir\'e TMDs, these energy scales are dependent on the moir\'e period $a_M$ (see Fig.~\ref{fig_Jandt}).
We accordingly present the holon results as functions of the $t/J$ (instead of $a_M$ in Fig.~\ref{fig_holon_bandwidth_dimensionless} and Fig.~\ref{fig_holon_hopping_coeff_dimensionless}) so that the results can be generalized to generic triangular lattices.
Similarly, we show how the exciton properties evolve with $t/J$ in Fig.~\ref{fig_exciton_properties_dimensionless}.
Note that here we fix the Coulomb energy scale $V_1$ while varying $t/J$ (whereas all energy scales change with $a_M$ for moir\'e TMDs.)

%% file: section_A_HP.tex
\section{Slave-fermion t-J model in the Holstein-Primakoff representation for spin}
\label{Appendix_Holstein_Primakoff}

In this section, we provide the details of Holstein-Primakoff (HP) representation of the slave fermion t-J model Eq.~\eqref{eq:slave_particle_H_t} and Eq.~\eqref{eq:slave_particle_H_J}.
Following Eq.~\eqref{eq:spinon_HP_boson_transform}, we express the spinon operator in terms of HP bosons as follows:
\begin{equation} 
\label{eq:HP_rep_spinon}
\hat{s}_{\bm{R},\tau}
=
\frac{1}{\sqrt{2}}
e^{i\tau\frac{2\pi\theta_{\bm{R}}}{3}}
\left(\sqrt{2S-\hat{a}_{\bm{R}}^\dagger\hat{a}_{\bm{R}}}-\tau\hat{a}_{\bm{R}}\right) 
,
\end{equation}
where $\theta_{\bm{R}}=0,1,-1$ for $A$, $B$ and $C$ sublattices, respectively, and $S$ is the magnitude of spin ($S=1/2$ in our problem). 
To simplify the problem and incorporate the depletion of magnetization by quantum fluctuation, we employ a mean-field approximation -- $\hat{a}_{\bm{R}}^\dagger\hat{a}_{\bm{R}}$ in Eq.~\eqref{eq:HP_rep_spinon} is replaced by $\frac{1}{N}\sum_{\bm{R}}\langle\hat{a}_{\bm{R}}^\dagger\hat{a}_{\bm{R}}\rangle$, where the expectation value is taken with respect to the mean field ground state of Eq.~\eqref{eq:slave_particle_H_J}.
This leads to:
\begin{equation} 
\label{eq:Quantum_Neel_order_ansatz}
\hat{s}_{\bm{R},\tau}
\to
\frac{1}{\sqrt{2}}
e^{i\tau\frac{2\pi\theta_{\bm{R}}}{3}}
\left(\xi-\tau\hat{a}_{\bm{R}}\right) 
,
\end{equation}
where $\xi=\sqrt{(1+m)/2}$ and $m$ is the sublattice magnetization in the HP representation:
\begin{equation}
m=1-\frac{2}{N}\sum_{\bm{R}}\langle\hat{a}_{\bm{R}}^\dagger\hat{a}_{\bm{R}}\rangle
.
\end{equation}
As a consequence of the transformation, the slave particle constraint Eq.~\eqref{eq:slave_particle_constraint} in the dilute charge limit, i.e. setting the holon occupation to zero, within the mean field approximation is expressed as:
\begin{equation}
\label{eq:constraint_HP_boson_MLSW}
\xi^2+\hat{a}_{\bm{R}}^\dag\hat{a}_{\bm{R}}=1
,
\end{equation}
and averaging out all moir\'e sites $\bm{R}$, it becomes:
\begin{equation}
\label{eq:avg_constraint_HP_boson_MLSW}
\xi^2+\frac{1}{N}\sum_{\bm{q}}\langle\hat{a}_{\bm{q}}^\dag\hat{a}_{\bm{q}}\rangle=1
.
\end{equation}
The slave fermion t-J model in the HP representation within the mean field approximation is derived as:
\begin{equation} \label{eq:mean_field_Hamiltonian}
\begin{aligned}
\hat{H}_{\textrm{tJ}}
&\approx
\hat{H}_t
+
\hat{H}_J
\\
\hat{H}_t
&=
\left(\frac{t\xi}{2}\right)
\sum_{\langle\bm{R},\bm{R}'\rangle}
\left[
\xi+\sqrt{3}\epsilon_{\bm{R}\bm{R}'}\hat{a}_{\bm{R}}
\right]
\hat{\psi}_{\bm{R}}^\dag
\hat{\psi}_{\bm{R}'}
+
h.c.
\\
\hat{H}_J
&=
\lambda\sum_{\bm{R}}
\hat{a}_{\bm{R}}^\dag\hat{a}_{\bm{R}}
\\
&
+
\left(\frac{J\xi^2}{8}\right)
\sum_{\langle\bm{R},\bm{R}'\rangle}
\left[
\hat{a}_{\bm{R}}^\dag\hat{a}_{\bm{R}'}
-3
\hat{a}_{\bm{R}}^\dag\hat{a}_{\bm{R}'}^\dag
+
h.c.
\right]
,
\end{aligned}
\end{equation}
where $\epsilon_{\bm{R}\bm{R}'}$ is the Levi-Civita symbol which is anti-symmetric, i.e. $\epsilon_{\bm{R}\bm{R}'}=-\epsilon_{\bm{R}'\bm{R}}$, and depends only on $\theta_{\bm{R}}$, the sublattice label of $\bm{R}$.
Explicitly, $\epsilon_{AB}=\epsilon_{BC}=\epsilon_{CA}=1$ and $\epsilon_{BA}=\epsilon_{CB}=\epsilon_{AC}=-1$.
The constraint in Eq.~\eqref{eq:avg_constraint_HP_boson_MLSW} can be incorporated through a Lagrangian multiplier $\lambda$, which takes the value of $3J\xi^2/2$, determined by by minimizing $\langle\hat{H}_J\rangle$.
Also, $\lambda=3J\xi^2/2$ gives gapless spin-wave excitations, consistent with the Goldstone mode from spontaneous symmetry breaking of continuous symmetry.
We also ignore the processes involving more than one HP boson, e.g. $\hat{\psi}^\dagger\hat{\psi}\hat{a}^\dagger\hat{a}$ and $\hat{a}^\dagger\hat{a}^\dagger\hat{a}\hat{a}$, since we are interested in the magnetic ordered state.
In the main text, Eq.~\eqref{eq:working_H_t} and Eq.~\eqref{eq:working_H_J} are derived with a Bogoliubov rotation $\hat{\beta}_{\bm{q}}=u_{\bm{q}}\hat{a}_{\bm{q}}-v_{\bm{q}}\hat{a}_{-\bm{q}}^\dag$, where $u_q$ and $v_q$ are defined in Eq.~\eqref{eq:Bogoliubov_u_coefficient} and Eq.~\eqref{eq:Bogoliubov_v_coefficient}, respectively.

To determine the equilibrium magnetization self-consistently, we apply the Bogoliubov rotation to Eq.~\eqref{eq:avg_constraint_HP_boson_MLSW}, giving:
\begin{equation}
\label{eq:self_consistency_eqn_m}
\begin{aligned}
\xi^2
&=
1-\frac{1}{N}\sum_{\bm{q}}\nu_{\bm{q}}^2
\\
m
&=
1-\frac{2}{N}\sum_{\bm{q}}v_{\bm{q}}^2
.
\end{aligned}
\end{equation}
We calculate the equilibrium magnetization by numerically doing the sum on the right hand side of Eq.~\eqref{eq:self_consistency_eqn_m} for system size as large as possible.
Here we do such calculation for system with size $N=3L^2$ for $L=500,1000,...,3000$, following with an extrapolation using a linear fitting between $m$ and $L^{-1}$.
It turns out that at $L\to\infty$, the result obtained is $m\simeq 0.47896$.
Hence, in our calculation we take $m=0.48$ as the equilibrium magnetization, which is close to the value reported in literature~\cite{Jolicoeur1989Spin}.

%% file: section_A_SCBA.tex
\section{Numerical solution of the dressed holon dispersion within SCBA} 
\label{Appendix_SCBA}

In this Appendix, we discuss the numerical procedures for solving SCBA. 
To numerically solve Eq.~\eqref{eq:SCBA_self_energy}, we use the fact that all $\omega_{\bm{q}}$ are non-negative.
We also exclude the momenta with $\omega_{\bm{q}}=0$ since the corresponding states do not belong to spin excitation.
Therefore, from Eq.~\eqref{eq:SCBA_self_energy}, we find that $\Sigma_{\bm{k}}(\epsilon)$ can be always expressed in terms of $\Sigma_{\bm{k}}(\epsilon')$ with $\epsilon'<\epsilon$.
With a sufficiently negative $\epsilon$ we can approximate the dressed holon self energy as
\begin{equation}
\label{eq:approximated_self_energy}
\Sigma_{\bm{k}}(\epsilon)
\approx
\frac{3t^2 \xi^2}{N}\sum_{\bm{q}} \frac{M_{\bm{k},\bm{q}}^2}{\epsilon-\omega_{\bm{q}}-t\xi^2\gamma_{\bm{k}+\bm{q}}}
.
\end{equation}
For numerical implementation, we pick an $\epsilon'<0$ that is large enough in magnitude such that $\Sigma_{\bm{k}}(\epsilon')$ satisfies Eq.~\eqref{eq:approximated_self_energy}.
Denoting small increment in $\epsilon$ as $\Delta\epsilon$, $\Sigma_{\bm{k}}(\epsilon'+\Delta\epsilon)$ is determined by $\Sigma_{\bm{k}}(\epsilon')$ according to Eq.~\eqref{eq:SCBA_self_energy}.
Hence, for $\epsilon-\epsilon'$ being multiples of $\Delta\epsilon$, we can generate $\Sigma_{\bm{k}}(\epsilon)$ recursively.

The dressed holon dispersion is determined by the pole of the dressed holon propagator $G_{\bm{k}}(\epsilon)$, which is given by
\begin{equation}
\label{eq:dressed_holon_propagator}
G_{\bm{k}}(\epsilon)=\frac{1}{\epsilon-t\xi^2\gamma_{\bm{k}}-\Sigma_{\bm{k}}(\epsilon)+i0_+}
,
\end{equation}
where the infinitesimal regulator $0_+$ is added as ($0.1t$) to implement the numerical calculation.
Hence, the dressed holon dispersion can be obtained by numerically solving $\epsilon_{\bm{k}}=\Sigma_{\bm{k}}(\epsilon_{\bm{k}})+t\xi^2\gamma_{\bm{k}}$.

A potential issue in numerically solving the self-consistent equations is that there might be mulitple solutions, and it is not guaranteed that the solution obtained in this way is the one of interest, which is the lowest energy one.
Here we alternatively solve for the dressed holon dispersion by finding the lowest energy peak of the spectral function $-\frac{1}{\pi}ImG_{\bm{k}}(\epsilon)$.

%% file: section_A_HF.tex
\section{Hartree-Fock analysis of the hole dispersion in triangular lattice Hubbard model}
\label{Appendix_Hartree_Fock}

In this section, we provide an alternative analysis to the hole dispersion for the triangular lattice Hubbard model from Eq.~\eqref{eq:two_body_Hamiltonian}.
We begin by considering a Hartree-Fock trial Hamiltonian including the hopping term of the Hubbard model and a sublattice Zeeman splitting field:
\begin{equation}
\label{eq:HF_trial_Hamiltonian}
\begin{aligned}
\hat{H}_0
&=
\hat{H}_t
+
\hat{H}_Z
\\
\hat{H}_Z
&=
-h_Z
\sum_{\tau,\tau'}
\sum_{\bm{R}}
\hat{h}_{\bm{R},\tau}
(
\hat{\bm{\sigma}}_{\tau\tau'}
\cdot
\hat{\bm{n}}_{\bm{R}}
)
\hat{h}_{\bm{R},\tau'}^\dagger
,
\end{aligned}
\end{equation}
where $\hat{\bm{n}}_{\bm{R}}$ are sublattice unit vectors defined in Eq.~\eqref{eq:Classical_Neel_order_ansatz}, and $h_Z$ is the variational parameter characterizing the strength of the sublattice Zeeman field in the trial Hamiltonian.
We use this Zeeman field term to capture the effect of $120^\circ$ spin order from the triangular lattice Hubbard model.
Note that this trial Hamiltonian is quadratic in fermion while the original Hubbard Hamiltonian is interacting.

Next, we obtain a trial density matrix $\hat{\rho}_0=Z_0^{-1}e^{-\frac{\hat{H}_0}{T}}$ from this trial Hamiltonian $\hat{H}_0$:
\begin{equation}
\label{eq:HF_free_energy}
F[h_Z]
=
\langle \hat{H}-\hat{H}_0\rangle_{\hat{\rho}_0}
+
F_0[h_Z]
,
\end{equation}
where we write the entropy term of the free energy as $T\langle\log\hat{\rho}_0\rangle_{\hat{\rho}_0}=F_0[h_Z]-\langle \hat{H}_0\rangle_{\hat{\rho}_0}$.
To determine $\hat{\rho}_0$, and hence $h_Z$, we minimize the free energy with respect to the variational parameter $h_Z$, giving:
\begin{equation}
\label{eq:HF_eqn_general}
\partial_{h_Z}\langle \hat{H}_U-\hat{H}_Z\rangle_{\hat{\rho}}
=
-\partial_{h_Z}F_0[\hat{\rho}]
,
\end{equation}
in which we set $\hat{H}$ to be the triangular lattice Hubbard model in Eq.~\eqref{eq:two_body_Hamiltonian}, with the on-site repulsion denoted as $\hat{H}_U$.
The expectation value $\langle\hat{H}_U\rangle$ is derived as:
\begin{equation}
\label{eq:HF_avg_HubbardU}
\langle
\hat{H}_U
\rangle_{\hat{\rho}_0}
=
\frac{NU}{3}
\sum_{\theta_{\bm{R}}}
\frac{1-\bm{m}(\theta_{\bm{R}})^2}{4}
,
\end{equation}
where $\theta_{\bm{R}}\in\{0,1,-1\}$ labels the sublattice of moir\'e site $\bm{R}$, as defined in Appendix~\ref{Appendix_Holstein_Primakoff}, and $\bm{m}(\theta_{\bm{R}})$ denotes the Hartree-Fock sublattice order parameter, which is expressed as:
\begin{align}
\bm{m}(\theta_{\bm{R}})
&=
\frac{3}{N}
\sum_{\tau,\tau'}
\sum_{\bm{R}\in\theta_{\bm{R}}}
\hat{\bm{\sigma}}_{\tau\tau'}
\langle
\hat{h}_{\bm{R},\tau}
\hat{h}_{\bm{R},\tau'}^\dagger
\rangle_{\hat{\rho}_0}
.
\end{align}
With these expressions, we reduce Eq.~\eqref{eq:HF_eqn_general} to
\begin{equation}
\label{eq:Hartree_Fock_equation}
h_Z
\hat{\bm{n}}_{\bm{R}}
=
\frac{U}{2}
\bm{m}(\theta_{\bm{R}})
.
\end{equation}

To continue, we need to diagonalize the trial Hamiltonian $\hat{H}_0$.
The convenient way to do this is to apply a spin rotation $\hat{U}_{\bm{R}}$ from $\bm{e}_z$ to $\hat{\bm{n}}_{\bm{R}}$, which is defined previously in Eq.~\eqref{eq:spinon_HP_boson_transform}, and define the rotated fermion operator $\hat{\tilde{h}}_{\bm{R},\tilde{\tau}}^\dagger=\sum_{\tau}[\hat{U}_{\bm{R}}]_{\tilde{\tau},\tau}\hat{h}_{\bm{R},\tau}^\dagger$, where $\tilde{\tau}=\{+,-\}$ labels the spin state aligned and anti-aligned to $\hat{\bm{n}}_{\bm{R}}$, respectively.
We take:
\begin{equation}
\label{eq:explicit_spin_rotation_matrix}
\hat{U}_{\bm{R}}
\equiv
\hat{U}(\theta_{\bm{R}})
=
\frac{1}{\sqrt{2}}
\begin{bmatrix}
e^{i\frac{2\pi}{3}\theta_{\bm{R}}}
&
e^{-i\frac{2\pi}{3}\theta_{\bm{R}}}
\\
-e^{i\frac{2\pi}{3}\theta_{\bm{R}}}
&
e^{-i\frac{2\pi}{3}\theta_{\bm{R}}}
\end{bmatrix}
.
\end{equation}
This makes the Hartree-Fock trial Hamiltonian become:
\begin{equation}
\begin{aligned}
\hat{H}_0
&=
\left(-\frac{t}{2}\right)
\sum_{\tilde{\tau}}
\sum_{\langle\bm{R},\bm{R}'\rangle} 
\hat{\tilde{h}}_{\bm{R},\tilde{\tau}}
\hat{\tilde{h}}_{\bm{R}',\tilde{\tau}}^\dagger
\\
&
+
\frac{i\sqrt{3}t}{2}
\sum_{\tilde{\tau},\tilde{\tau}'}
\sum_{\langle\bm{R},\bm{R}'\rangle}
\epsilon_{\bm{R}\bm{R}'}
\hat{\tilde{h}}_{\bm{R},\tilde{\tau}}
(
\hat{\bm{\sigma}}_{\tilde{\tau},\tilde{\tau}'}
\cdot
\bm{e}_x
)
\hat{\tilde{h}}_{\bm{R}',\tilde{\tau}'}^\dagger
\\
&
-h_Z
\sum_{\tilde{\tau},\tilde{\tau}'}
\sum_{\bm{R}}
\hat{\tilde{h}}_{\bm{R},\tilde{\tau}}
(
\hat{\bm{\sigma}}_{\tilde{\tau},\tilde{\tau}'}
\cdot
\bm{e}_z
)
\hat{\tilde{h}}_{\bm{R},\tilde{\tau}'}^\dagger
,
\end{aligned}
\end{equation}
where $\epsilon_{\bm{R}\bm{R}'}$ is the anti-symmetric tensor defined in Appendix~\ref{Appendix_Holstein_Primakoff}.
In momentum space representation, we have:
\begin{equation}
\begin{aligned}
\hat{H}_0
&=
\sum_{\tau,\tau'}
\sum_{\bm{k}}
\hat{\tilde{h}}_{\bm{k},\tau}
\bigg[
-t
\gamma_{\bm{k}}
-
h_Z
(\hat{\bm{\sigma}}_{\tau\tau'}
\cdot\bm{e}_z)
\\
&
-
\sqrt{3}t
h_{\bm{k}}
(\hat{\bm{\sigma}}_{\tau\tau'}
\cdot\bm{e}_x)
\bigg]
\hat{\tilde{h}}_{\bm{k},\tau'}^\dagger
,
\end{aligned}
\end{equation}
with $h_k$ defined in Eq.~\eqref{eq:h_coefficient}.
The spectrum of $\hat{H}_0$ follows directly as $\epsilon_{k,\zeta}=-t\gamma_k+\zeta\Upsilon_k$ with the Hartree-Fock bands labeled by $\zeta=\pm1$. 
The energy splitting is given by:
\begin{equation}
\Upsilon_{\bm{k}}
=
\sqrt{h_Z^2+3t^2h_{\bm{k}}^2}
.
\end{equation}
The eigen-modes $\hat{\tilde{h}}_{\bm{k},\zeta}$ are described by $\hat{\tilde{h}}_{\bm{k},\tau}^\dagger=\sum_\zeta W_{\tau\zeta}(\bm{k})\hat{\tilde{h}}_{\bm{k},\zeta}^\dagger$ with the following transformation coefficients:
\begin{align}
W_{\tau\zeta}(\bm{k})
=
(\zeta sgn[h_{\bm{k}}])^{\frac{1-\tau}{2}}
\sqrt{\frac{1}{2}\left[1-\tau\zeta\frac{h_Z}{\Upsilon_{\bm{k}}}\right]}
.
\end{align}
These relations simplfies Eq.~\eqref{eq:Hartree_Fock_equation} to:
\begin{equation}
\label{eq:Hartree_Fock_equation_final}
h_Z
=
\frac{U}{2N}
\sum_{\bm{k}}
\frac{h_Z}{\Upsilon_{\bm{k}}}
,
\end{equation}
which allows for determination of $h_Z$.
As $U\gg t$, we have $h_Z\simeq\frac{U}{2}$.

The dressed holon dispersion discussed in the main text should correspond to a particle-hole transformation to the $\zeta=-1$ band in the context of Hartree-Fock calculation, which gives:
\begin{equation} \label{eq:Hartree_Fock_dispersion}
\begin{aligned}
\epsilon_k^{\textrm{(HF)}}
&=
t\gamma_k
+
\frac{U}{2}\sqrt{1+12\left(\frac{th_k}{U}\right)^2}
\\
&
\simeq
\frac{U}{2}
+
\left(t-\frac{3J}{2}\right)
\gamma_k
+
\frac{3J}{4}
\tilde{\gamma}_k
-
\frac{3J}{4}
\gamma_{2k}
\end{aligned}
\end{equation}
in the limit $U\gg t$.
This dispersion is equivalent to Eq.~\eqref{eq:approximate_single_particle_dispersion} up to an overall constant, and we identify that $2t_1=-t+3J/2$, $2t_2=-3J/4$ and $2t_3=3J/4$.

%% file: section_A_Mottmoire_current.tex
\section{Current operator and Optical conductivity for Mott-moir\'e exciton} 
\label{Appendix_slave_fermion_mean_field_current}

In this section, we discuss the derivation of current operator Eq.~\eqref{eq:Current_operator} in Section~\ref{sec:formalism}.

To start with, the low-energy physics for Bloch electrons near the valleys within monolayer TMDs is modeled by the massive Dirac fermion model~\cite{Xiao2012}.
Hence, we describe twisted TMD bilayer as a monolayer system experiencing the moir\'e potentials from the other layer, meaning that the low-energy physics of valley $\tau$ and layer $l$ is given by:
\begin{equation}
\label{eq:Massive_Dirac_model}
\hat{H}_{l,\tau}
=
\begin{bmatrix}
E_l^g/2 & v_l^F(\tau\hat{p}_{l,x}-i\hat{p}_{l,y}) \\
v_l^F(\tau\hat{p}_{l,x}+i\hat{p}_{l,y}) & -E_l^g/2
\end{bmatrix}
+
\Delta_l(\bm{r}_l)
+
\hat{H}_{SO}
,
\end{equation}
where the basis states for the matrix are the orbitals of electrons in the conduction and valence bands, respectively.
$v_l^F$, $E_l^g$ and $\Delta_l(\bm{r}_l)$ are the Fermi velocity, the band gap and the moir\'e potential of layer $l$, respectively.
$\hat{\bm{p}}_l$ denotes the momentum operator measured from the valleys and $\bm{r}_l$ is the position variable for each layer.
Note that $\bm{r}_l$ is discretized by $a_l$, the lattice spacing of layer $l$, it is often treated as continuous since $a_l$ is much smaller than the periodicity of the moir\'e potential $a_M$ for small twist angles~\cite{Rivera2018Interlayer}. 
$\hat{H}_{SO}$ denotes the spin-orbit coupling term that is responsible for spin-valley locking of low energy degrees of freedom in TMD.

The moir\'e length scale $a_M$ splits the Brillouin zone for the monolayers into small mBZs.
Hence, it is sufficient to consider $\bm{p}_l$ within the first mBZ such that Eq.~\eqref{eq:Massive_Dirac_model} reduces to the decoupled moir\'e-Hamiltonians for the valence and conduction bands, which are folded by the moir\'e potential into VmB and CmB, respectively.
Focusing on the lowest CmB and the highest VmB, which can be described by tight-binding/Hubbard models~\cite{Wu2018Hubbard,Li2021Imaging}, we see that Eq.~\eqref{eq:Massive_Dirac_model} becomes Eq.~\eqref{eq:Starting_model}.

To derive the expression Eq.~\eqref{eq:Current_operator}, we replace the momentum operator with $\hat{\bm{p}}_l\to\hat{\bm{p}}_l+(e/c)\bm{A}$, where $c$ is the speed of light and $\bm{A}$ is the vector potential, and take the functional derivative of the Hamiltonian Eq.~\eqref{eq:Massive_Dirac_model} with respective to $\bm{A}$. 
Next, we outline the derivation from Eq.~\eqref{eq:Current_operator} to the optical conductivity Eq.~\eqref{eq:Optical_conductivity_Mott}.
We consider LSW to linearize Eq.~\eqref{eq:HP_rep_spinon} for simplicity, and ignore all spin fluctuations in the current.
In other words, we consider only the classical 120$^{\circ}$ spin-ordered state for the spin sector, since this dominates in the light-matter coupling $\hat{\bm{j}}\cdot\bm{A}$, as long as the strength of the vector potential $\bm{A}$ is small.

From Eq.~\eqref{eq:HP_rep_spinon}, within LSW treatment, the slave fermion substitution Eq.~\eqref{eq:slave_particle_def} for the electron operator in VmB follows as:
\begin{equation}
\label{eq:slave_fermion_MF_classical_Neel}
\hat{h}_{\bm{R},\tau}^\dagger
=
\frac{1}{\sqrt{2}}
e^{i\tau\frac{2\pi\theta_{\bm{R}}}{3}}
\left(1-\tau\hat{a}_{\bm{R}}\right)
\hat{\psi}_{\bm{R}}^\dagger
\end{equation}
with $\theta_{\bm{R}}=\{0,1,-1\}$ for the three sublattices as defined in Appendix~\ref{Appendix_Holstein_Primakoff}. 
The current operator Eq.~\eqref{eq:Current_operator} then becomes:
\begin{equation}
\label{eq:current_slave_fermion}
\hat{\bm{j}}^{(cv)}
\simeq
\frac{ev_F}{c}
\sum_{\bm{k},\tau}
\bm{e}_\tau
\hat{c}_{\bm{k},\tau}^\dagger
\hat{\psi}_{-\bm{k}-\tau\bm{\kappa}}^\dagger
+
h.c.
,
\end{equation}
in which we neglect the spin-fluctuation term $\hat{c}^\dagger\hat{\psi}^\dagger\hat{a}$, as mentioned previously, and we use the properties of the sublattice plane-wave factor, $e^{i\tau\frac{2\pi\theta_{\bm{R}}}{3}}=e^{-i\tau\bm{\kappa}\cdot\bm{R}}$ with $\bm{\kappa}=\frac{4\pi}{3}\bm{e}_x$ the momentum labeling $\kappa$ in mBZ (in units of $a_M^{-1}$), assuming the origin $\bm{R}=0$ takes $\theta_{\bm{R}}=0$.
We proceed to rewrite Eq.~\eqref{eq:current_slave_fermion} in terms of exciton operator Eq.~\eqref{eq:Exciton_operator}:
\begin{equation}
\label{eq:current_exciton}
\hat{\bm{j}}^{(cv)}
\simeq
\frac{\sqrt{N}ev_F}{c}
\sum_{n,\tau}
\bm{e}_\tau
\Phi_{-\tau\bm{\kappa}}^{(n)}
\hat{X}_{n,\tau}(-\tau\bm{\kappa})
+
h.c.
,
\end{equation}
in which we use $\Phi_{\bm{Q}}^{(n)}=\frac{1}{\sqrt{N}}\sum_{\bm{p}}\phi_{\bm{Q}}^{(n)}(\bm{p})$. 
The optical matrix element~\cite{Wu2018Theory} for the Mott-moir\'e exciton state $\hat{X}_{n,\tau}(\bm{Q})$, denoted as $\bm{J}_{n,\tau}(\bm{Q})$, is then:
\begin{equation}
\label{eq:Optical_matrix_element_Mott-moire}
\begin{aligned}
\bm{J}_{n,\tau}(\bm{Q})
&
\equiv
\frac{1}{\sqrt{{\cal{A}}}}
\langle GS|
\hat{\bm{j}}^{(cv)}
\hat{X}_{n,\tau}^\dagger(\bm{Q})
|GS\rangle
\\
&
=
\sqrt{\frac{N}{{\cal{A}}}}
\frac{ev_F}{c}
\delta_{\bm{Q},-\tau\bm{\kappa}}
\bm{e}_\tau
\Phi_{-\tau\bm{\kappa}}^{(n)}
,
\end{aligned}
\end{equation}
where ${\cal{A}}$ denotes the system area such that ${\cal{A}}/N$ gives the area of unit moir\'e cell, and $|GS\rangle$ is the exciton ground state.
The optical conductivity from linear response theory~\cite{Marel2004Optical} follows as $\sigma_{ij}(\omega)=\sigma(\omega)\delta_{ij}$, where
\begin{equation}
\label{eq:Optical_conductivity_Kubo}
\sigma(\omega)
\simeq
\frac{i}{\omega}
\sum_{n,\tau,\bm{Q}}
\frac{|\bm{J}_{n,\tau}(\bm{Q})|^2}{\omega-E_{n,\bm{Q}}^X+i\eta}
,
\end{equation}
where we neglect the branch with $\omega+E_{n,\bm{Q}}^X+i\eta$ since we are interested near resonance, i.e. $\omega\simeq E_{n,\bm{Q}}^X$.
Putting Eq.~\eqref{eq:Optical_matrix_element_Mott-moire} into Eq.~\eqref{eq:Optical_conductivity_Kubo}, we arrive at Eq.~\eqref{eq:Optical_conductivity_Mott}.

%% file: section_A_moire_conductivity.tex
\section{Optical conductivity for moir\'e exciton} 
\label{Appendix_moire_exciton_optical_conductivity}

For moir\'e excitons (i.e., the absence of intraband correlation),
the interband current operator at zero momentum is:
\begin{align}
\hat{\bm{j}}^{(cv)}
=
\frac{\sqrt{2}ev_F}{c}
\sum_{\bm{k},\tau}
\hat{e}_\tau
\hat{c}_{\bm{k},\tau}^\dagger
\hat{h}_{\bm{k},\tau}^\dagger
+h.c.
\end{align}
and with similar to Eq.~\eqref{eq:Exciton_operator} for Mott-moir\'e exciton, we define the moire exciton operator as:
\begin{equation}
\begin{aligned}
\hat{X}_{n,\tau}(\bm{Q})
=
\sum_{\bm{p}}
\phi_{\bm{Q}}^{(n)}(\bm{p})
\hat{h}_{-\frac{\bm{Q}}{2}+\bm{p},\tau}
\hat{c}_{\frac{\bm{Q}}{2}+\bm{p},\tau}
,
\end{aligned}
\end{equation}
where $\phi$ in this section denotes the moire exciton wavefunction.
The current from moire exciton follows as:
\begin{align}
\bm{j}^{(cv)}
=
\frac{\sqrt{2N}ev_F}{c}
\sum_{n,\tau}
\hat{e}_\tau
\Phi_0^{(n)}
\hat{X}_{n,\tau}^\dagger
+h.c.
,
\end{align}
where $\hat{X}_{n,\tau}^\dagger\equiv\hat{X}_{n,\tau}^\dagger(0)$ and $\Phi_0^{(n)}=\frac{1}{\sqrt{N}}\sum_{\bm{p}}
\phi_0^{(n)}(p)$.
Similar to the calculation presented in Appendix~\ref{Appendix_slave_fermion_mean_field_current}, we obtain the optical conductivity for moir\'e exciton as $\sigma_{ij}(\omega)=\sigma(\omega)\delta_{ij}$ with:
\begin{equation}
\label{eq:Optical_conductivity_moire}
\sigma(\omega)
\sim
\frac{i}{\omega}
\sum_{n,\tau}
\frac{|\Phi_0^{(n)}|^2}{\omega-E_{n,0}^X+i0_+}
,
\end{equation}
where in this section $E_{n,0}^X$ denotes the energy of moire exciton at zero total momentum.

%% file: section_A_perturb.tex
\section{Perturbation theory on Wannier equation for excitons in TMD heterobilayer}
\label{Appendix_perturbation}

\begin{figure}[t]
\centering
\includegraphics[width=1.0\columnwidth]{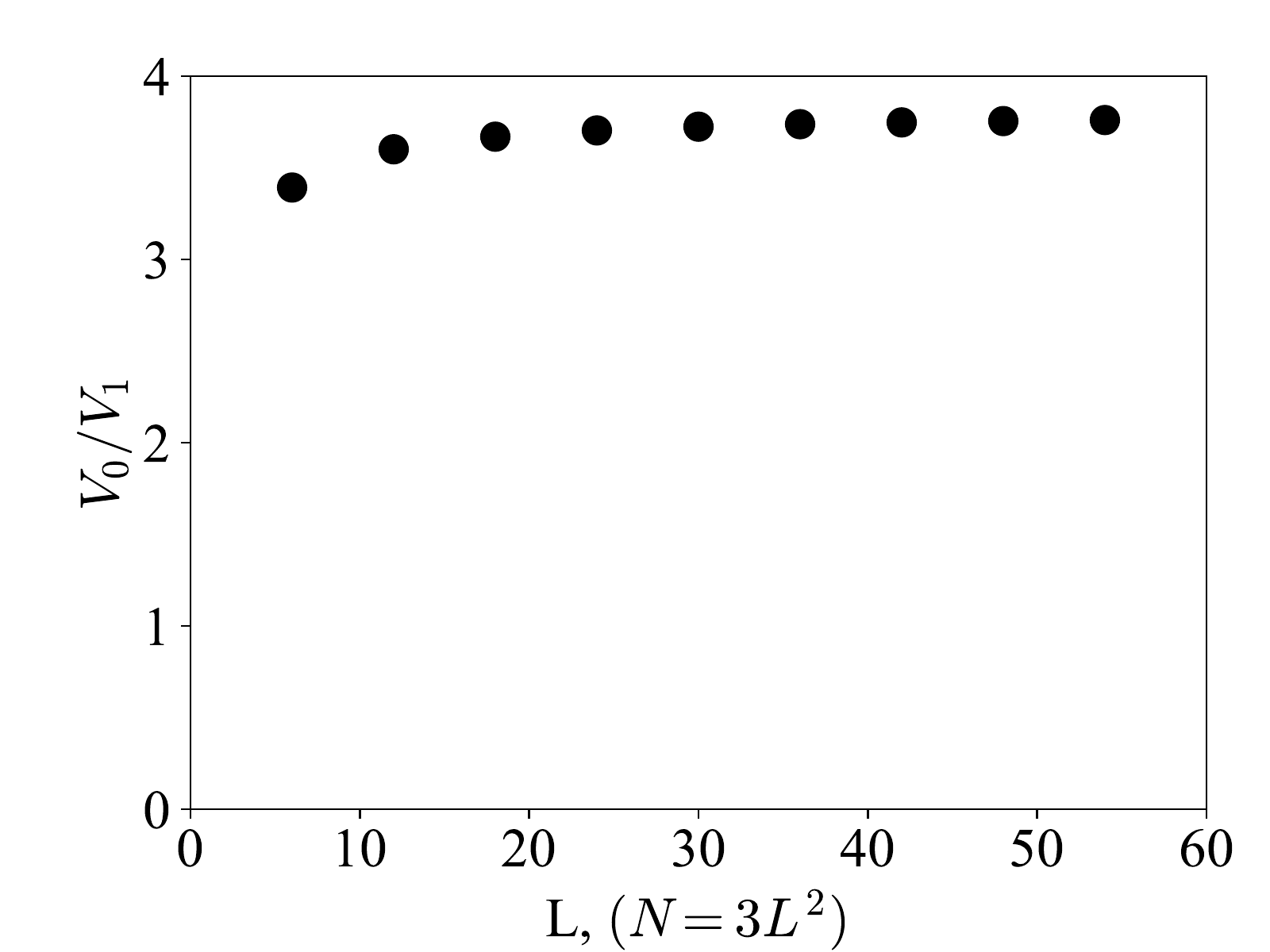}
\caption{Ratio between $V_0$ and $V_1$ from Eq.~\eqref{eq:Coulomb_gate_screening} for different system sizes $N=3L^2$.
As $\bm{q}$ is summed over mBZ, $V_0/V_1$ asymptotically approaches a specific value.
}
\label{fig_V0_L}
\end{figure}

\begin{figure}[t]
\centering
\includegraphics[width=1.0\columnwidth]{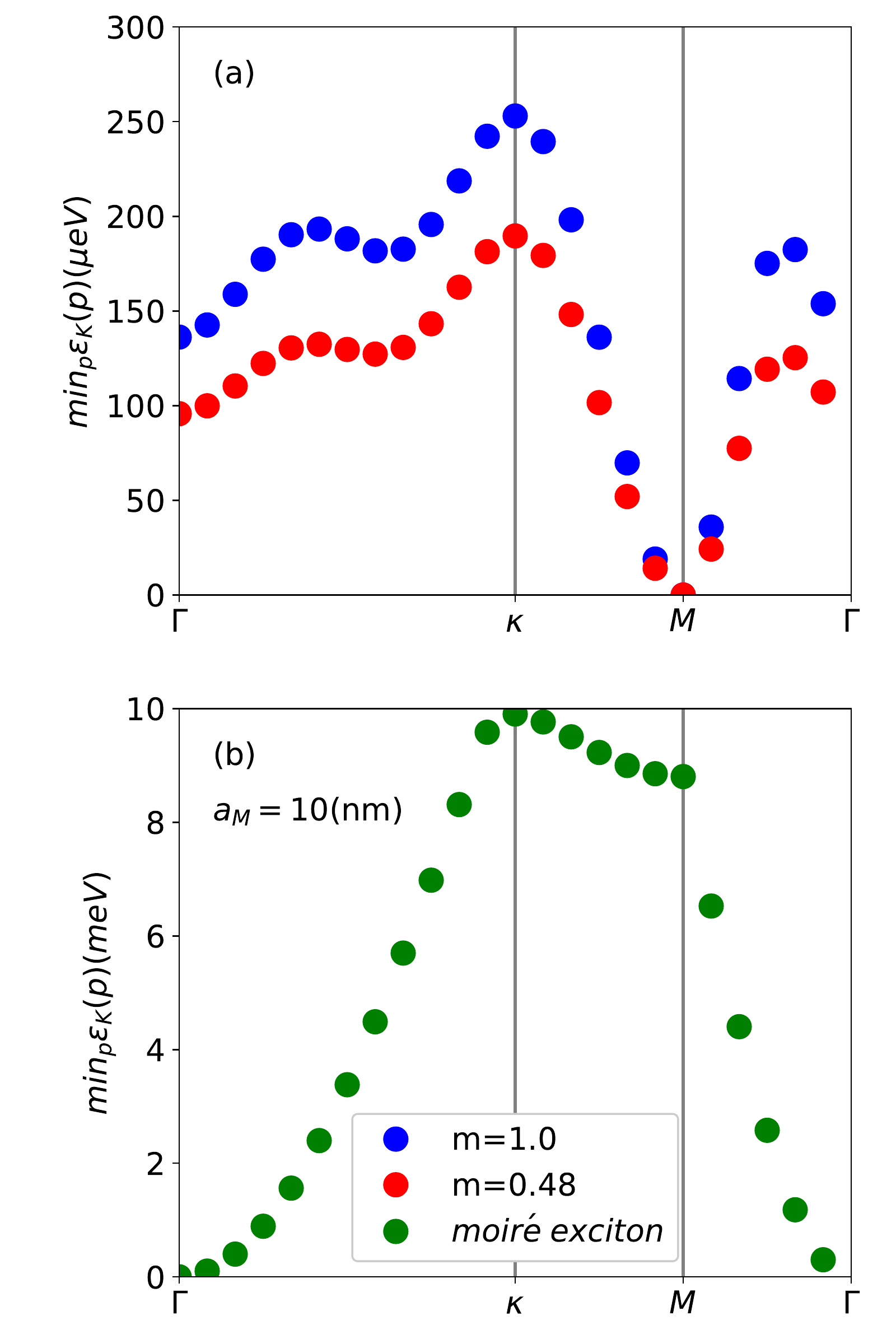}
\caption{Center-of-mass motion for the lowest internal states for the free two-particle system, i.e. not affected by mutual Coulomb attraction, of a lattice of size $3\times 24 \times 24$ in the case with (top) and without (bottom) Mott physics. 
The moir\'e period is set as $a_M=10$(nm).
In the case without Mott physics, the hole motion is described by $\hat{H}_v\to (-2t)\sum_{\bm{k}}\gamma_{\bm{k}}\hat{h}_{\bm{k},\tau}^\dagger\hat{h}_{\bm{k},\tau}$ in Eq.~\eqref{eq:two_body_Hamiltonian}.
}
\label{fig_unbound_dispersion}
\end{figure}

In this section, we discuss the perturbation theory on Eq.~\eqref{eq:Wannier_equation} in detail.
First, we point out that the total momentum $\bm{Q}$ is a good quantum number of the Hamiltonian operator in Eq.~\eqref{eq:Wannier_equation}, meaning that the energy eigen-states are also eigen-states of $\bm{Q}$.
Thus, these energy eigen-states can be labeled as $|n,\bm{Q}\rangle$, with $n$ labeling the internal states.
We suppress the valley degeneracy throughout the discussion in this section.

We consider the strong interacting limit for perturbation, i.e. $\langle\varepsilon_{\bm{Q}}(\bm{p})\rangle_{n,\bm{Q}}\ll\langle V(q)\rangle_{n,\bm{Q}}$.
We also emphasize that such a limit is of interest in our work as the exciton dispersion is flat compared to the Coulomb binding for both Mott-moir\'e and moir\'e excitons, as indicated by Fig.~\ref{fig_exciton_properties}.
The unperturbed term in the Hamiltonian for Eq.~\eqref{eq:Wannier_equation} is then the Coulomb attraction term $V(q)$, which gives unperturbed states as eigenstates of the relative distance operator $\hat{r}$ according to the position space representation of $V(q)$.
Hence, we have $|n,\bm{Q}\rangle\simeq|j,\bm{Q}\rangle$ in the strong interacting limit, where $j$ are non-negative integers that label $|\bm{r}|$ in non-descending order.

The unperturbed ground state is the state with $|\bm{r}|=0$, denoted as $|0,\bm{Q}\rangle$.
We denote the unperturbed energy for this state as $-V_0$, which we estimate to be ${\cal{A}}^{-1}\sum_{\bm{q}}V(q)\simeq3.7V_1$ with $\bm{q}$ summed over mBZ and $N=3\times 24^2$ (see Fig.~\ref{fig_V0_L}), since we are assuming the electrons and holes are tightly-bound to moir\'e sites.
We expect this estimation to capture the correct qualitative properties with this perturbation scheme, since we have $V_0\gg t$ using the above expression.
In reality, we expect a smaller $V_0$ due to the finite width of Wannier functions of the quasi-particles, but we expect $V_0$ to be of order $U$ in this case and hence $V_0\gg t$ is still valid.

The first-order correction on ground state energy $\delta E_{0,\bm{Q}}^{X,(1)}=\langle\varepsilon_{\bm{Q}}(\bm{p})\rangle_{0,\bm{Q}}$ is zero.
This comes from the fact that the momentum space wavefunction of the unperturbed ground state $\langle p,\bm{Q}|0,\bm{Q}\rangle=N^{-1/2}$ is just a constant, and that $\varepsilon_{\bm{Q}}(\bm{p})$ is composed of sinusoidal functions, for both Mott-moir\'e and moir\'e excitons.
Hence, the leading order correction to the ground state energy is at least second order. 

Calculations of the second order correction $\delta E_{0,\bm{Q}}^{X,(2)}$ require the information of unperturbed excited states.
The first few unperturbed excited states are labeled as $|1,\bm{Q}\rangle$, $|2,\bm{Q}\rangle$ and $|3,\bm{Q}\rangle$, which has well-defined relative distance $|\bm{r}|=1,\sqrt{3}$ and $2$ (in units of $a_M$), respectively.
We denote the corresponding energies as $-V_{1,2,3}$, of which magnitude are much smaller than $V_0$.
Note that we suppress the label for the six-fold degeneracy for each $|j>0,\bm{Q}\rangle$ for simplicity.
We then proceed for the second order correction $\delta E_{0,\bm{Q}}^{X,(2)}$, which involves matrix elements $\langle j,\bm{Q}|\varepsilon_{\bm{Q}}(\bm{p})|0,\bm{Q}\rangle$ for $j>0$. 

We start with the perturbation for moir\'e excitons.
For moir\'e excitons, only $j=1$ contributes since the position space representation of $\varepsilon_{\bm{Q}}(\bm{p})$ contains only nearest neighboring hopping terms.
We then obtain
\begin{align}
\langle 1,\bm{Q}|\varepsilon_{\bm{Q}}(\bm{p})|0,\bm{Q}\rangle
=
(-2t)\cos\left(\pm\frac{\bm{Q}}{2}\cdot\bm{e}_i\right)
,
\end{align}
where $\bm{e}_i$ are nearest neighboring vectors defined below Eq.~\eqref{eq:Bogoliubov_v_coefficient}.
The second order correction for moir\'e excitons follows as:
\begin{equation}
\label{eq:pertubation_moire_exciton_dispersion}
\delta E_{0,\bm{Q}}^{X,(2)}
=
-\frac{4t^2}{V_0-V_1}\gamma_{\bm{Q}}
-\frac{12t^2}{V_0-V_1}
.
\end{equation}

The situation is slightly more complicated for Mott-moir\'e excitons, in which terms with $j=1,2,3$ would contribute to $\delta E_{0,\bm{Q}}^{X,(2)}$.
Nevertheless, only the $j=1$ term contribute to the exciton bandwidth $W_X$.
Direct evaluation gives:
\begin{align}
\langle 1,\bm{Q}|\varepsilon_{\bm{Q}}(\bm{p})|0,\bm{Q}\rangle
&=
(-t-t_1)\cos\left(\pm\frac{\bm{Q}}{2}\cdot\bm{e}_i\right)
\nonumber\\
&
+i(t-t_1)\sin\left(\pm\frac{\bm{Q}}{2}\cdot\bm{e}_i\right)
,
\end{align}
\begin{align}
\langle 2,\bm{Q}|\varepsilon_{\bm{Q}}(\bm{p})|0,\bm{Q}\rangle
=
(-2t_2)e^{i\frac{\bm{Q}}{2}\cdot\bm{r}_2}
,
\end{align}
\begin{align}
\langle 3,\bm{Q}|\varepsilon_{\bm{Q}}(\bm{p})|0,\bm{Q}\rangle
=
(-t_3)e^{i\frac{\bm{Q}}{2}\cdot\bm{r}_3}
,
\end{align}
where $\bm{r}_{2,3}$ denotes the relative separation for states with $j=2,3$, respectively.
Consequently, the second order correction for Mott-moir\'e excitons is:
\begin{equation}
\label{eq:pertubation_Mott_moire_exciton_dispersion}
\begin{aligned}
\delta E_{0,\bm{Q}}^{X,(2)}
&=
-\frac{6(t^2+t_1^2)}{V_0-V_1}
-\frac{6t_2^2}{V_0-V_2}
-\frac{6t_3^2}{V_0-V_3}
\\
&
-\frac{4tt_1}{V_0-V_1}\gamma_{\bm{Q}}
.
\end{aligned}
\end{equation}
Comparing the results for moir\'e exciton and Mott-moir\'e exciton, the ratio between their bandwidths is:
\begin{equation}
\label{eq:bandwidth_ratio_perturbation}
\frac{W_X^{Mm}}{W_X^{m}}
=
\left|\frac{t_1}{t}\right|\ll1
,
\end{equation}
where $W_X^{Mm}$ and $W_X^{m}$ denotes the bandwidths of lowest Mott-moir\'e exciton and moir\'e exciton, respectively. 
Comparison between numerical and perturbation results is shown in Fig.~\ref{fig_bandwidth_perturbation}. 


Next, we continue to use this perturbative analysis to investigate the exciton binding energy.
We start from the qualitative observation that correction to exciton energy $E_{0,\bm{Q}}^X$ is at most of order $t^2/V_0$ for both Mott-moir\'e and moir\'e excitons.
Hence, up to linear order in $t$, we can approximate $E_{0,\bm{Q}}^X\simeq -V_0-\mu$, where $\mu$ denotes the chemical potential that is set differently for the two types of excitons.
Recall that we define the exciton binding energy as the energy reduction from the lowest-branch unbound two particle kinetic energy to the exciton energy, i.e. $E_{0,\bm{Q}}^B\equiv\min_{\bm{p}}\varepsilon_{\bm{Q}}(\bm{p})-E_{0,\bm{Q}}^X$ with $\varepsilon_{\bm{Q}}(\bm{p})$ as the unbound two particle kinetic energy defined in Eq.~\eqref{eq:Two_particle_kinetic_energy}.
This definition reflects that the Coulomb binding conserves the total momentum $\bm{Q}$.
An example of $\min_{\bm{p}}\varepsilon_{\bm{Q}}(\bm{p})$ is plotted in Fig.~\ref{fig_unbound_dispersion}, suggesting that the width of $\min_{\bm{p}}\varepsilon_{\bm{Q}}(\bm{p})$ in $\bm{Q}$ is of order $J$ for Mott-moir\'e exciton and of order $t$ for moir\'e exciton.
From direct calculations, we find that to the linear order in $t$, $\min_{\bm{p}}\varepsilon_{\bm{Q}}(\bm{p})$ is $-6t-\mu$ for Mott-moir\'e exciton and $-2t\gamma_{\bm{Q}}-6t-\mu$ for moir\'e exciton.
Hence, to the linear order in $t$, the Mott-moir\'e exciton binding energy is $E_{0,\bm{Q}}^B\simeq V_0-6t$, while for moir\'e exciton it is $E_{0,\bm{Q}}^B\simeq V_0-6t-2t\gamma_{\bm{Q}}$, which is $V_0-3t$ at $\bm{Q}=\bm{\kappa}$.
This explains the slightly larger binding for moir\'e exciton, as illustrated in Fig.~\ref{fig_exciton_properties}(a).
We end by pointing out that the non-negligible dependence of binding energy on total momentum $\bm{Q}$ for moir\'e exciton is very different from the case for hydrogenic exciton~\cite{Haug2004}.
This is because the center of mass degrees of freedom can be separated from the relative motion for hydrogenic exciton, while these degrees of freedoms are not separable for excitons derived from the moir\'e superlattice~\cite{Mattis1986The}.